\documentclass{article}
\usepackage{graphicx}
\usepackage{amssymb}
\begin{document}

\title{Optimisation of Quantum Hamiltonian Evolution:\\
       From Two Projection Operators to Local Hamiltonians}
\author{Apoorva Patel$^*$ and Anjani Priyadarsini$^\dagger$\\
        Centre for High Energy Physics, Indian Institute of Science,\\
        Bangalore 560012, Karnataka, India\\
        $^*$adpatel@cts.iisc.ernet.in\\
        $^\dagger$anjanipriyav@cts.iisc.ernet.in\\}

\date{}
\maketitle

\begin{abstract}
Given a quantum Hamiltonian and its evolution time, the corresponding
unitary evolution operator can be constructed in many different ways,
corresponding to different trajectories between the desired end-points
and different series expansions. A choice among these possibilities can
then be made to obtain the best computational complexity and control
over errors. It is shown how a construction based on Grover's algorithm
scales linearly in time and logarithmically in the error bound, and is
exponentially superior in error complexity to the scheme based on the
straightforward application of the Lie-Trotter formula. The strategy is
then extended first to simulation of any Hamiltonian that is a linear
combination of two projection operators, and then to any local efficiently
computable Hamiltonian. The key feature is to construct an evolution in
terms of the largest possible steps instead of taking small time steps.
Reflection operations and Chebyshev expansions are used to efficiently
control the total error on the overall evolution, without worrying about
discretisation errors for individual steps. We also use a digital
implementation of quantum states that makes linear algebra operations
rather simple to perform.
\end{abstract}


\section{Introduction}

Richard Feynman advocated development of quantum computers as efficient
simulators of physical quantum systems \cite{feynman}. Real physical
systems are often replaced by simplified models in order to understand
their dynamics. Even then, exact solutions are frequently not available,
and it has become commonplace to study the models using elaborate computer
simulations. Classical computer simulations of quantum models are not
efficient---well-known examples range from the Hubbard model to lattice
QCD---and Feynman argued that quantum simulations would do far better.

As a concrete realisation of Feynman's argument, it is convenient to look
at Hamiltonian evolution of a many-body quantum system. Quantum simulations
can sum multiple evolutionary paths contributing to a quantum process in
superposition at one go, while classical simulations need to evaluate these
paths one by one. Formalisation of this advantage, in terms of computational
complexity, has gradually improved over the years. Real physical systems
are governed by local Hamiltonians, i.e. where each component interacts only
with a limited number of its neighbours independent of the overall size of
the system. Lloyd constructed a quantum evolution algorithm for such systems
\cite{lloyd}, based on the discrete time Lie-Trotter decomposition of the
unitary evolution operator, and showed that it is efficient in the required
time and space resources. Aharonov and Ta-Shma rephrased the problem as
quantum state generation, treating the terms in the Hamiltonian as black box
oracles, and extended the result to sparse Hamiltonians in graph theoretical
language \cite{aharonov}. The time complexity of the algorithm was then
improved \cite{BACS,WBHS,childs}, using Suzuki's higher order generalisations
of the Lie-Trotter formula \cite{suzuki} and clever decompositions of the
Hamiltonian. Recently the error complexity of the evolution has been reduced
from power-law to logarithmic in the inverse error, using the strategy of
discrete time simulation of multi-query problems \cite{BCCKS1}. This is a
significant jump in computational complexity improvement that needs
elaboration and understanding. In this article, we explicitly construct
efficient evolution algorithms first for Hamiltonians that are linear
combinations of two projection operators, \emph{to expose the physical
reasons behind the improvement}, and then extend the strategy to any local
efficiently computable Hamiltonian. Our constructive methods differ from
the reductionist approach of Ref.~\cite{BCCKS1}, we improve upon earlier
results, and clearly demonstrate how the algorithms work in practice.

Computational complexity of a problem is a measure of the resources
needed to solve it. Conventionally, the computational complexity of a
decision problem is specified in terms of the size of its input, noting
that the size of its output is only one bit. This framework is extended
to problems with different output requirements (e.g. find the optimal route
for the travelling salesman problem or evaluate $\pi$ to a certain precision),
by setting up successive verifiable bounds on the outputs. For example,
the problem of evaluating $\pi$ can be implemented as first confirming
that $\pi\in[3,4]$, and then narrowing down the interval by bisection,
adding one bit of precision for every decision made. In such a scenario,
the number of decision problems solved equals the number of output bits,
and the complexity of the original problem is the sum of the complexities
for the individual decision problems. It is therefore appropriate to
specify the complexity of the original problem in terms of the size of its
input as well as its output, especially for function evaluation problems.
Generalizing the conventional classification, \emph{the computational
algorithm can then be labeled efficient if the required resources are
polynomial in terms of the size of both its input and its output}. We label
such algorithms as belonging to the class P:P, explicitly expressing their
computational complexity with respect to their input as well as output
sizes. Simultaneous consideration of both input and output dependence of
complexity is natural for reversible computation. It is also necessary
when extending finite precision analog computation to arbitrary precision
digital computation. Note that our definition of P:P efficiency differs
from the concept of ``simulatable Hamiltonians" in Ref.~\cite{aharonov}.

The traditional computational complexity analysis (e.g. P vs. NP
classification) does not discuss much how the complexity depends on the
output precision, and the task is relegated to design of efficient methods
for arbitrary precision numerical analysis \cite{brent}. We stress that
both input and output size dependence of the computational complexity
are equally important for practical function evaluation problems.
Popular importance sampling methods are not efficient according to our
criterion, because the number of iterations needed in the computational
effort has a negative power-law dependence on the precision $\epsilon$
(i.e. $N_{\rm iter}\propto\epsilon^{-2}$ as per the central limit theorem).
On the other hand, finding zeroes of a function by bisection is efficient
(i.e. $N_{\rm iter}\propto\log\epsilon$), and finding them by Newton's
method is super-efficient (i.e. $N_{\rm iter}\propto\log\log\epsilon$).

We describe the Hamiltonian simulation problem in Section \ref{qhamsim},
along with the important ingredients required for its optimisation.
In Section \ref{qdshamevol}, we formulate the database search problem as
Hamiltonian evolution. While the evolution results are well-known (see
for instance Refs.~\cite{prevcomp1,prevcomp2}), we focus on the error
complexity which has not been optimised in the literature. Our analysis
explicitly shows how Grover's large time-step algorithm is exponentially
superior to small time-step algorithms approximating continuous time
evolution. It also demonstrates that the large time-step algorithm
effectively simulates a very different Hamiltonian than the small time-step
algorithm, but yields the same total evolution operator \cite{LAT2014}.
As an important ingredient, we introduce digital representation for
quantum states that makes performing linear algebra operations with them
straightforward. In Section \ref{projreflser}, we construct a series
expansion evolution algorithm for Hamiltonians that are linear combinations
of two projection operators. We carry out a partial summation of the series,
and demonstrate how evaluation of a truncated series of large-step reflection
operators improves the error complexity exponentially compared to the
small-step Lie-Trotter formula. Our analytic results are supported by
numerical tests.  Finally in Section \ref{localham}, we combine the methods
of Chebyshev series expansion and digital representation, to construct
an efficient simulation algorithm for any local efficiently computable
Hamiltonian. We conclude with an outlook for our methods, and some general
results for projection operators are collected in an Appendix.

\section{Quantum Hamiltonian Simulation}
\label{qhamsim}

The Hamiltonian simulation problem is to evolve an initial quantum state
$|\psi(0)\rangle$ to a final quantum state $|\psi(T)\rangle$, in presence
of interactions specified by a Hamiltonian $H(t)$:
\begin{equation}
\label{unitevol}
|\psi(T)\rangle = U(T) |\psi(0)\rangle ~,~~
U(T) =  {\cal P} \Big[ \exp\big(-i\int_0^T H(t) dt\big) \Big] ~.
\end{equation}
The initial state can often be prepared easily, while the final state
is generally unknown. The path ordering of the unitary evolution
operator $U(T)$, denoted by the symbol ${\cal P}$ in Eq.(\ref{unitevol}),
is necessary when various terms in the Hamiltonian do not commute.
The properties of the final state are subsequently obtained from
expectation values of various observables:
\begin{equation}
\label{expectobs}
\langle O_a \rangle =  \langle\psi(T)|O_a|\psi(T)\rangle ~.
\end{equation}
In typical problems of quantum dynamics, both these parts---the final
state and the expectation values---are determined probabilistically
upto a specified tolerance level. They also require different techniques,
and so it is convenient to deal with them separately. In this article,
we focus only on the former part; the latter part has been addressed in
Refs.~\cite{harrow,clader}, and still needs exponential improvement in
the dependence of computational complexity on the output precision to belong
to the class P:P. For simplicity, we also restrict ourselves to problems
where both the Hamiltonian $H$ and the observables $O_a$ are bounded.%
\footnote{Physical problems with unbounded Hamiltonians and
          operators exist---the Coulomb interaction is a well-known case.
          Their numerical solutions need more sophisticated techniques.}

It is also possible to define the Hamiltonian simulation problem as the
determination of the evolution operator $U(T)$, and omit any mention
of the initial and the final states. The accuracy of the simulation is
then specified by the norm of the difference between simulated and exact
evolution operators, say $||\widetilde{U}(T)-U(T)|| < \epsilon$. In actual
implementation, the simulated $\widetilde{U}(T)$ may not be exactly unitary,
due to round-off and truncation errors, but the preceding measure for the
accuracy of the simulation still suffices as long as $\epsilon$ is small
enough.

We concern ourselves here only with Hamiltonians acting in finite
$N$-dimensional Hilbert spaces. A general Hamiltonian would then be a
dense $N \times N$ matrix, and there is no efficient way to simulate it.
So we restrict the Hamiltonian according to the following features
commonly present in physical problems:\\
(1) The Hilbert space is a tensor product of many small components,
e.g. $N=2^n$ for a system of $n$ qubits.\\
(2) The components have only local interactions irrespective of the
size of the system, e.g. only nearest neighbour couplings. That makes
the Hamiltonian sparse, with $O(N)$ non-zero elements.\\
(3) The Hamiltonian is specified in terms of a finite number of efficiently
computable functions, while the arguments of the functions can depend on
the components, e.g. the interactions are translationally invariant.\\
These features follow the notion of Kolmogorov complexity, where the
computational resources needed to describe an object are quantified in
terms of the compactness of the description. With a compact description
of the Hamiltonian, the resources needed to just write it down do not
influence the simulation complexity.%
\footnote{Even Hamiltonians without explicit symmetry structure can have
          a compact description with a compressed labeling scheme, as in
          case of finite element domain decompositions. Mathematically,
          ``local interactions" can be traded for ``limited interactions",
          and the number of functions can be enlarged somewhat, but such
          possibilities are unlikely in common physical problems.}

Such sparse Hamiltonians can be mapped to graphs with bounded degree $d$,
with the vertices representing the physical components of the system and
the edges denoting the interactions between neighbouring components.
Their simulations can be easily parallelised---on classical computers,
Hamiltonians with these features allow SIMD simulations with domain
decomposition.

We note that long range physical interactions do exist, but simulation of
generic dense Hamiltonians is not efficient \cite{BerryChilds}. Only with
some extra properties, dense Hamiltonians can lead to non-local evolution
operators having compact descriptions. A useful example is FFT, which
describes a dense but factorisable unitary transformation that can be
efficiently implemented, but we do not consider such possibilities here.

With all these specifications, \emph{efficient Hamiltonian simulation
algorithms in the class P:P use computational resources that are polynomial
in $\log(N)$, $d$ and $\log(\epsilon)$}.

\subsection{Hamiltonian Decomposition}
\label{hamdecomp}

Efficient simulation strategy for Hamiltonian evolution has two major
ingredients. In general, exponential of a sparse Hamiltonian is not
sparse, which makes exact evaluation of $\exp(-iHt)$ difficult.
So the first ingredient is to decompose the sparse Hamiltonian as
a sum of non-commuting but block-diagonal Hermitian operators, i.e.
$H =\sum_{i=1}^l H_i$. The motivation for such a decomposition is twofold:

\noindent
(a) Functions of individual $H_i$, defined as power series, can be easily
and exactly calculated for any time evolution $\tau$, and they retain the
same block-diagonal structure.

\noindent
(b) The blocks are decoupled and so can be evolved simultaneously,
in parallel (classically) or in superposition (quantum mechanically).

\noindent
Furthermore, the blocks can be reduced in size all the way to a mixture
of $1\times1$ and $2\times2$ blocks. The $1\times1$ blocks just produce
phases upon exponentiation, while the $2\times2$ blocks can be expressed
as linear combinations of identity and projection or reflection operators
(i.e. $(1+\hat{n}\cdot\vec{\sigma})/2$ or $\hat{n}\cdot\vec{\sigma}$
respectively, where $\hat n$ is a unit vector and $\sigma_i$ are the
three Pauli matrices). There is no loss of generality in such a choice;
it is just a convenient choice of basis that simplifies the subsequent
algorithm. Projection or reflection operators with only two distinct
eigenvalues can be interpreted as binary query oracles. Their large
spectral gaps also help in rapid convergence of series expansions
involving them.

% Optimal edge colouring for bipartite graphs is $O(n_{edge} \log(d)$.
% Generic optimal edge colouring problem is NP-complete.

In general, $H_i$ can be systematically identified by an edge-colouring
algorithm for graphs \cite{aharonov}, with distinct colours (labeled by
the index $i$) for overlapping edges. As per Vizing's theorem, any simple
graph of degree $d$ can be efficiently coloured with $d+1$ colours.
Physical models are often defined on bipartite graphs, for which the
colouring algorithms are simpler than those for general graphs and need
$d$ colours. Identification of $H_i$ also provides a compressed labeling
scheme that can be used to address individual blocks.

Actual calculations do not need explicit construction of $U(T)$, rather
only the effect of $U(T)$ on the quantum state $|\psi(0)\rangle$ has to
be evaluated. That is accomplished by breaking down the calculation into
steps, each of which consists of the product of a sparse matrix with a
vector, e.g. $\exp(-iH_i\tau)|\psi\rangle$. The simulation complexity is
then conveniently counted in units of such sparse matrix-vector products.

As a simple illustration, the discretised Laplacian for a one-dimensional
lattice has the block-diagonal decomposition given by:
\begin{eqnarray}
\label{oneDlap}
& & \left(\matrix{
    \cdots&\cdots  &\cdots  &\cdots  &\cdots  &\cdots  &\cdots \cr
    \cdots&\hfill-1&\hfill 2&\hfill-1&\hfill 0&\hfill 0&\cdots \cr
    \cdots&\hfill 0&\hfill-1&\hfill 2&\hfill-1&\hfill 0&\cdots \cr
    \cdots&\hfill 0&\hfill 0&\hfill-1&\hfill 2&\hfill-1&\cdots \cr
    \cdots&\cdots  &\cdots  &\cdots  &\cdots  &\cdots  &\cdots \cr
}\right) \nonumber\\
&=& \left(\matrix{
    \cdots&\cdots  &\cdots  &\cdots  &\cdots  &\cdots  &\cdots \cr
    \cdots&\hfill-1&\hfill 1&\hfill 0&\hfill 0&\hfill 0&\cdots \cr
    \cdots&\hfill 0&\hfill 0&\hfill 1&\hfill-1&\hfill 0&\cdots \cr
    \cdots&\hfill 0&\hfill 0&\hfill-1&\hfill 1&\hfill 0&\cdots \cr
    \cdots&\cdots  &\cdots  &\cdots  &\cdots  &\cdots  &\cdots \cr
}\right)
 +  \left(\matrix{
    \cdots&\cdots  &\cdots  &\cdots  &\cdots  &\cdots  &\cdots \cr
    \cdots&\hfill 0&\hfill 1&\hfill-1&\hfill 0&\hfill 0&\cdots \cr
    \cdots&\hfill 0&\hfill-1&\hfill 1&\hfill 0&\hfill 0&\cdots \cr
    \cdots&\hfill 0&\hfill 0&\hfill 0&\hfill 1&\hfill-1&\cdots \cr
    \cdots&\cdots  &\cdots  &\cdots  &\cdots  &\cdots  &\cdots \cr
}\right) .
\end{eqnarray}
This decomposition, $H=H_o+H_e$, has the projection operator structure
following from $H_o^2=2H_o$ and $H_e^2=2H_e$. Graphically, the break-up
can be represented as:
\begin{center}
\setlength{\unitlength}{2pt}
\begin{picture}(120,10)
\thicklines
\put(10,5){\line(1,0){20}}  \put(18,7){\makebox(0,0)[bl]{o}}
\put(50,5){\line(1,0){20}}  \put(58,7){\makebox(0,0)[bl]{o}}
\put(90,5){\line(1,0){20}}  \put(98,7){\makebox(0,0)[bl]{o}}
\put(10,5){\circle*{2}}     \put(30,5){\circle*{2}}
\put(50,5){\circle*{2}}     \put(70,5){\circle*{2}}
\put(90,5){\circle*{2}}     \put(110,5){\circle*{2}}
        
\thinlines
\put(30,5){\line(1,0){20}}  \put(38,7){\makebox(0,0)[bl]{e}}
\put(70,5){\line(1,0){20}}  \put(78,7){\makebox(0,0)[bl]{e}}

\put(0,5){\makebox(0,0)[bl]{$\ldots$}}
\put(113,5){\makebox(0,0)[bl]{$\ldots$}}
\end{picture}
\end{center}
where $H_o$ and $H_e$ are identified by the last bit of the position label.
Eigenvalues of $H$ are $4\sin^2(k/2)$ in terms of the lattice momentum $k$,
while those of $H_o$ and $H_e$ are just $0$ and $2$.

\subsection{Evolution Optimisation}

Given that individual $H_i$ can be exponentiated exactly and efficiently,
their sum $H$ can be approximately exponentiated using the discrete
Lie-Trotter formula:
\begin{eqnarray}
\label{LieTrotter}
\exp\big(-i H T\big) &=& \exp\Big(-i\sum_i H_i T\Big) \\
  &\approx& \Big(\prod_i \exp(-iH_i \Delta t)\Big)^m ~,~~ m = T/\Delta t ~.
  \nonumber
\end{eqnarray}
This replacement maintains unitarity of the evolution exactly, but may
not preserve other properties such as the energy. The accuracy of the
approximation is commonly improved by making $\Delta t$ sufficiently small,
sometimes accompanied by higher order discretisations.%
\footnote{Time-dependent Hamiltonians are expanded about the
          mid-point of the interval $\Delta t$ for higher accuracy.}
This approach has been used for classical parallel computer simulations
of quantum evolution problems \cite{deraedt,richardson}.

In contrast, the second ingredient of efficient Hamiltonian simulation is
to use as large $\Delta t$ as possible. When the exponent is proportional
to a projection operator, the largest $\Delta t$ is the one that makes
the exponential a reflection operator.%
\footnote{On the unitary sphere, the farthest one can move from an initial
          state along a specified direction is to the diametrically opposite
          state, and that is the reflection operation.}
In general, use of any fixed constant $\Delta t$ changes the leading scaling
behaviour of the error complexity from a power-law dependence on $\epsilon$
to a logarithmic one. The extreme strategy of choosing the largest possible
$\Delta t$ not only keeps the evolution accurate by reducing the round-off
and the truncation errors, but also optimises the scaling proportionality
constant.%
\footnote{For example, Grover's algorithm has query complexity
          $(\pi/4)\sqrt{N}$. The leading $\sqrt{N}$ scaling can be achieved
          using operators with $\Theta(1)$ phase shifts, while optimisation
          of the scaling coefficient to $\pi/4$ is achieved using reflection
          operators corresponding to phase shifts equal to $\pi$.}
It is not at all obvious how such a result may arise, and so we demonstrate
it first in Section 3 using the database search problem as an explicit
example, and then in Section 4 for Hamiltonians that are a linear combination
of two more general projection operators.

Our efficient Hamiltonian simulation algorithms, explicitly constructed
using the two ingredients just described, have computational complexity
\begin{equation}
O\left( t \frac{\log(t/\epsilon)}{\log(\log(t/\epsilon))} {\cal C} \right) ~.
\end{equation}
Here ${\cal C}$ is the computational cost of a single time step,
which only weakly depends on $t$ and $\epsilon$. It is ${\cal C}$ that
characterises how computational complexity of classical implementation is
improved in the quantum case, through conversion of independent parallel
execution threads into quantum superposition. We point out that to keep
the discretisation error under control, digital calculations need $b$-bit
precision, with $b=\Omega(\log((t/\epsilon)\log(t/\epsilon)))$. For
$l$-sparse Hamiltonians whose elements can be evaluated efficiently,
the computational cost ${\cal C}$ is $O(lNb^3)$ classically and $O(lnb^3)$
quantum mechanically.

\section{Quantum Database Search as Hamiltonian Evolution}
\label{qdshamevol}

The quantum database search algorithm works in an $N$-dimensional Hilbert
space, whose basis vectors are identified with the individual items.
It takes an initial state whose amplitudes are uniformly distributed
over all the items, to the target state where all but one amplitudes vanish.
Let $\{|i\rangle\}$ be the set of basis vectors, $|s\rangle$ be the
initial uniform superposition state, and $|t\rangle$ be the target state
corresponding to the desired item. Then
\begin{eqnarray}
&& |\psi(0)\rangle = |s\rangle ~,~~ |\psi(T)\rangle = |t\rangle ~, \nonumber\\
&& |\langle i|s \rangle| = 1/\sqrt{N} ~,~~ \langle i|t \rangle = \delta_{it} ~.
\end{eqnarray}

The simplest evolution schemes taking $|s\rangle$ to $|t\rangle$ are
governed by time-independent Hamiltonians that depend only on $|s\rangle$
and $|t\rangle$. The unitary evolution is then a rotation in the
two-dimensional subspace, formed by $|s\rangle$ and $|t\rangle$,
of the whole Hilbert space. In this subspace, let
\begin{equation}
|t\rangle = \begin{pmatrix} 1 \\ 0 \end{pmatrix} ~,~~
|t_\perp\rangle = \begin{pmatrix} 0 \\ 1 \end{pmatrix} ~,~~
|s\rangle = \begin{pmatrix} 1/\sqrt{N} \\ \sqrt{(N-1)/N} \end{pmatrix} ~.
\end{equation}
On the Bloch sphere representing the density matrix, the states
$|s\rangle\langle s|$ and $|t\rangle\langle t|$ are respectively given by
the unit vectors $\hat{n}_s=\big(\frac{2\sqrt{N-1}}{N},0,\frac{2}{N}-1\big)$
and $\hat{n}_t=(0,0,1)$. The angle between them is $\cos^{-1}(\frac{2}{N}-1)$,
which is twice the angle $\cos^{-1}(1/\sqrt{N})$ between $|s\rangle$
and $|t\rangle$ in the Hilbert space.

For a time-independent Hamiltonian, the time evolution of the state is a
rotation at a fixed rate around a direction specified by the Hamiltonian:
\begin{equation}
U(t) = \exp(-iHt) = \exp(-i\hat{n}_H\cdot\vec{\sigma}~\omega t) ~.
\end{equation}
For the database search problem, $U(T)|s\rangle=|t\rangle$, upto a phase
arising from a global additive constant in the Hamiltonian. There are many
possible evolution routes from the initial to the target state, and we
consider two particular cases in turn.

\subsection{Farhi-Gutmann's and Grover's Algorithms}

Grover based his algorithm on a physical intuition \cite{trotter},
where the potential energy term in the Hamiltonian attracts the wavefunction
towards the target state and the kinetic energy term in the Hamiltonian
diffuses the wavefunction over the whole Hilbert space. Both the potential
energy $|t\rangle\langle t|$ and the kinetic energy $|s\rangle\langle s|$%
\footnote{It is the mean field version of kinetic energy corresponding
          to the maximally connected graph.}
terms are projection operators. The corresponding time-independent
Hamiltonian is
\begin{eqnarray}
\label{HamFG}
H_C  =  |s\rangle\langle s| + |t\rangle\langle t|
    &=& \pmatrix{ 1+\frac{1}{N}        & \frac{\sqrt{N-1}}{N} \cr
                  \frac{\sqrt{N-1}}{N} & 1-\frac{1}{N} } \\
    &=& I + \frac{\sqrt{N-1}}{N} \sigma_1 + \frac{1}{N} \sigma_3 ~. \nonumber
\end{eqnarray}
That gives rise to the evolution operator (omitting the global phase)
\begin{eqnarray}
U_C(t) &=& \exp\big(-i\hat{n}\cdot\vec{\sigma}~t/\sqrt{N}\big) \nonumber\\
       &=& \cos(t/\sqrt{N}) - i\hat{n}\cdot\vec{\sigma}\sin(t/\sqrt{N}) ~,
\end{eqnarray}
which is a rotation on the Bloch sphere by angle $2t/\sqrt{N}$ around the
direction $\hat{n} = \big(\sqrt{(N-1)/N}, 0, 1/\sqrt{N}\big)^T$.

The (unnormalised) eigenvectors of $H_C$ are $|s\rangle\pm|t\rangle$,
which are the orthogonal states left invariant by the evolution operator
$U_C(t)$. On the Bloch sphere, their density matrices point in the
directions $\pm\hat{n}$, which bisect the initial and the target states,
$|s\rangle\langle s|$ and $|t\rangle\langle t|$. Thus a rotation by angle
$\pi$ around the direction $\hat{n}$ takes $|s\rangle\langle s|$ to
$|t\rangle\langle t|$ on the Bloch sphere. In the Hilbert space,
the rotation angle taking $|s\rangle$ to $|t\rangle$ is then $\pi/2$,
and so the time required for the Hamiltonian search is $T=(\pi/2)\sqrt{N}$
\cite{farhi}.

Grover made an enlightened jump from this scenario, motivated by the
Lie-Trotter formula. He exponentiated the projection operators in $H_C$
to reflection operators; $R=\exp(\pm i\pi P)=1-2P$ for any projection
operator $P$. The optimal algorithm that Grover discovered iterates the
discrete evolution operator \cite{grover},
\begin{eqnarray}
\label{operG}
U_G &=& -(1-2|s\rangle\langle s|)(1-2|t\rangle\langle t|) \nonumber\\
    &=& \pmatrix{ 1-\frac{2}{N}          & 2\frac{\sqrt{N-1}}{N} \cr
                  -2\frac{\sqrt{N-1}}{N} & 1-\frac{2}{N}         }
     =  (1-\frac{2}{N}) I + 2i\frac{\sqrt{N-1}}{N} \sigma_2 ~.
\end{eqnarray}
With $U_G = \exp(-iH_G\tau)$, it corresponds to the evolution Hamiltonian
\begin{eqnarray}
H_G &=& \frac{i}{\sqrt{N}} \big(|t\rangle\langle s|-|s\rangle\langle t|\big)
    \nonumber\\
    &=& \pmatrix{ 0                      & i\frac{\sqrt{N-1}}{N} \cr
                  -i\frac{\sqrt{N-1}}{N} & 0                     }
     =  -\frac{\sqrt{N-1}}{N} \sigma_2 ~,
\end{eqnarray}
and the evolution step
\begin{equation}
\label{stepG}
\tau = \frac{N}{\sqrt{N-1}} \sin^{-1}\Big( 2\frac{\sqrt{N-1}}{N} \Big)
     = \frac{2N}{\sqrt{N-1}} \sin^{-1}\Big( \frac{1}{\sqrt{N}} \Big) ~.
\end{equation}
It is an important non-trivial fact that $H_G$ is the commutator
of the two projection operators in $H_C$:
\begin{equation}
H_G = i\big[|t\rangle\langle t|,|s\rangle\langle s|\big] ~.
\end{equation}
This commutator is the leading correction to the Lie-Trotter formula 
in the Baker-Campbell-Hausdorff (BCH) expansion \cite{BCHreview},
making Grover's algorithm an ingenious summation of the BCH expansion
for the evolution operator.

On the Bloch sphere, each $U_G$ step is a rotation by angle
$2\tau\sqrt{N-1}/N = 4\sin^{-1}(1/\sqrt{N})$ around the direction
$\hat{n}_G=(0,1,0)^T$, taking the geodesic route from the initial
to the final state. That makes the number of steps required for this
discrete Hamiltonian search,
\begin{eqnarray}
Q_T &=& \frac{1}{4} \cos^{-1}\Big( \frac{2}{N}-1 \Big) \Big/
                \sin^{-1}\Big( \frac{1}{\sqrt{N}} \Big) \nonumber\\
    &=& \frac{\cos^{-1}(1/\sqrt{N})}{2\sin^{-1}(1/\sqrt{N})}
    \approx \frac{\pi}{4}\sqrt{N} ~.
\end{eqnarray}

\begin{figure}[t]
\begin{center}
\includegraphics[width=6cm]{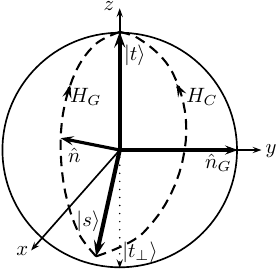}
\end{center}
\caption{Evolution trajectories on the Bloch sphere for the quantum
database search problem, going from $|s\rangle$ to $|t\rangle$.
The Hamiltonians $H_C$ and $H_G$ generate rotations around the
directions $\hat{n}$ and $\hat{n}_G$ respectively.}
\label{evoltraj}
\end{figure}

Note that $\hat{n}$ and $\hat{n}_G$ are orthogonal, so the evolution
trajectories produced by rotations around them are completely different
from each other, as illustrated in Fig.\ref{evoltraj}. It is only after
a specific evolution time, corresponding to the solution of the database
search problem, that the two trajectories meet each other.%
\footnote{Incidentally, the adiabatic quantum search algorithm,
          described by the time-dependent Hamiltonian
          $H(u) = (1-u)|s\rangle\langle s| + u|t\rangle\langle t|$
          with $u(t)\in[0,1]$, follows the same evolution trajectory as
          Grover's algorithm.}
The evolution operators in the two cases are:
\begin{equation}
U_C(T) = \exp\Big(-i\hat{n}\cdot\sigma \frac{\pi}{2}\Big)
       = -i \pmatrix{ \frac{1}{\sqrt{N}}  & \sqrt{\frac{N-1}{N}} \cr
                      \sqrt{\frac{N-1}{N}} & -\frac{1}{\sqrt{N}} \cr } ~,
\end{equation}
\begin{equation}
(U_G)^{Q_T} = \exp\Big(i\cos^{-1}\Big(\frac{1}{\sqrt{N}}\Big)\sigma_2\Big)
            = \pmatrix{ \frac{1}{\sqrt{N}}    & \sqrt{\frac{N-1}{N}} \cr
                        -\sqrt{\frac{N-1}{N}} & \frac{1}{\sqrt{N}}   \cr } ~.
\end{equation}

The rates of different Hamiltonian evolutions can be compared only after
finding a common convention to fix the magnitude and the ease of simulation
of the Hamiltonians:

\noindent
(1) Magnitude: A global additive shift of the Hamiltonian has no practical
consequence. So possible comparison criteria can be the norm of the traceless
part of the Hamiltonian or the spectral gap over the ground state.
$||H_C||=1/\sqrt{N}$ and $||H_G||=\frac{\sqrt{N-1}}{N}$ are comparable,
with the same limit as $N\rightarrow\infty$.

\noindent
(2) Ease of simulation: The binary query oracle can be used to produce
various functions of $|t\rangle\langle t|$.

\begin{figure}[t]
{
\begin{center}
%\setlength{\unitlength}{1mm}
% Unit length is 1pt
\begin{picture}(250,120)
\put(20,100){\line(1,0){20}}
\put(70,100){\line(1,0){70}}
\put(170,100){\line(1,0){20}}
\multiput(20,80)(5,0){4}{\line(1,0){2}}
\multiput(70,80)(5,0){14}{\line(1,0){2}}
\multiput(170,80)(5,0){4}{\line(1,0){2}}
\put(20,60){\line(1,0){20}}
\put(70,60){\line(1,0){70}}
\put(170,60){\line(1,0){20}}
\put(40,50){\framebox(30,60){Oracle}}
\put(140,50){\framebox(30,60){Oracle}}

\put(0,78){$|x\rangle$}
\put(12,90){$\bigg\lmoustache$} \put(12,66){$\bigg\rmoustache$}
\put(190,90){$\bigg\rmoustache$} \put(190,66){$\bigg\lmoustache$}
\put(200,78){$O_\phi|x\rangle$}

\put(55,16){\line(0,1){34}}
\put(155,16){\line(0,1){34}}
\put(55,20){\circle{8}}
\put(155,20){\circle{8}}

\put(20,20){\line(1,0){60}}
\put(130,20){\line(1,0){60}}
\put(80,0){\framebox(50,40){$\pmatrix{ 1 & 0 \cr 0 & e^{i\phi} }$}}
\put(5,18){$|0\rangle$}
\put(195,18){$|0\rangle$}
\end{picture}
\end{center}
}
\caption{Quantum logic circuit for the fractional query oracle operator
$O_\phi=\exp(i\phi|t\rangle\langle t|)$. The oracle flips the ancilla
bit iff its input is the target state, and the standard binary query
oracle operator corresponds to $\phi=\pi$.}
\label{fracoracle}
\end{figure}
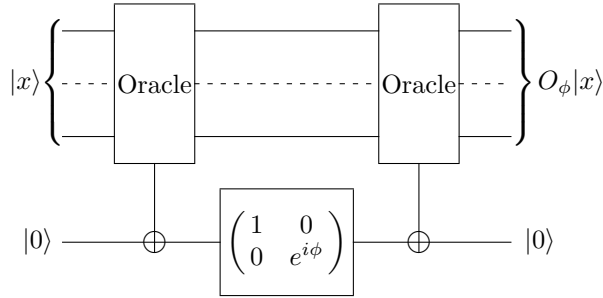

{\narrower\noindent
(a) $H_C$ can be easily simulated by alternating small evolution steps
governed by $|s\rangle\langle s|$ and $|t\rangle\langle t|$, according
to the Lie-Trotter formula \cite{nielsen}. Each evolution step governed
by $|t\rangle\langle t|$ needs two binary query oracles, as shown in
Fig.\ref{fracoracle}.

\noindent
(b) $U_G$ is easily obtained using one binary query oracle per evolution
step.

}

\subsection{Equivalent Hamiltonian Evolutions}

When different evolution Hamiltonians exist, corresponding to different
evolution routes from the initial to the final states, one can select an
optimal one from them based on their computational complexity and stability
property. This feature can be used to simplify the Hamiltonian evolution
problem by replacing the given Hamiltonian by a simpler equivalent one.
Two Hamiltonian evolutions are truly equivalent, when their corresponding
unitary evolution operators are the same (for a fixed evolution time and
upto a global phase). The intersection of the two evolution trajectories
is then independent of the specific initial and final states.

For the database search problem, we observe that
\begin{equation}
\label{equivevolint}
U_C(T) = i (1-2|t\rangle\langle t|) ~ (U_G)^{Q_T} ~.
\end{equation}
So with an additional binary query oracle, one evolution can be used as
an alternative for the other, without worrying about the specific choices
of $|s\rangle$ and $|t\rangle$. (The additional oracle is needed to make
the two Hamiltonian evolutions match, although it is not required for the
database search problem.)

For a more general evolution time $0<t<T$, we have the relation
(analogous to Euler angle decomposition),
\begin{equation}
\label{equivevolfrac}
U_C(t) = \exp\big(i\beta\sigma_3\big) ~(U_G)^{Q_t}~
         \exp\Big(i\big(\frac{\pi}{2}+\beta\big)\sigma_3\Big) ~,
\end{equation}
i.e. $U_C(t)$ can be generated as $Q_t$ iterations of the Grover
operator $U_G$, preceded and followed by phase rotations. Since
$\sigma_3=2|t\rangle\langle t|-1$, each phase rotation is a fractional
query oracle and can be obtained using two oracle calls \cite{nielsen}.
The parameters in Eq.(\ref{equivevolfrac}) are given by
\begin{eqnarray}
\label{evolfracparam}
Q_t &=& \frac{\sin^{-1}\Big(\sqrt{\frac{N-1}{N}} \sin\frac{t}{\sqrt{N}}\Big)}
             {2\sin^{-1}(1/\sqrt{N})} \approx \frac{t}{2} ~, \nonumber\\
\beta &=& -\frac{\pi}{4} -\frac{1}{2}
             \tan^{-1}\Big(\frac{1}{\sqrt{N}}\tan\frac{t}{\sqrt{N}}\Big) ~.
\end{eqnarray}
They yield
\begin{eqnarray}
& & (U_G)^{Q_t} = \exp\Big(i \sin^{-1}\Big(\sqrt{\frac{N-1}{N}}
                           \sin\frac{t}{\sqrt{N}}\Big) \sigma_2\Big) \\
&=& \pmatrix{ \sqrt{\cos^2\frac{t}{\sqrt{N}}
    + \frac{1}{N}\sin^2\frac{t}{\sqrt{N}}}
    & \sqrt{\frac{N-1}{N}}\sin\frac{t}{\sqrt{N}} \cr
     -\sqrt{\frac{N-1}{N}}\sin\frac{t}{\sqrt{N}}
    & \sqrt{\cos^2\frac{t}{\sqrt{N}}
    + \frac{1}{N}\sin^2\frac{t}{\sqrt{N}}} \cr } , \nonumber
\end{eqnarray}
whose elements are the same as those of $U_C(t)$ upto phase factors.

Thus $H_G$ can be used to obtain the same evolution as $H_C$, \emph{even
though the two Hamiltonians are entirely different in terms of their
eigenvectors and eigenvalues---a rare physical coincidence indeed!}
A straightforward conversion scheme is to break up the duration of
evolution for $H_C$ into units that individually solve a database search
problem, simulate each integral unit according to Eq.(\ref{equivevolint}),
and the remaining fractional part according to Eq.(\ref{equivevolfrac}).

\subsection{Unequal Magnitude Evolution Operators}

Now consider the generalisation of $H_C$ to the situation where the
coefficients of $|s\rangle\langle s|$ and $|t\rangle\langle t|$ are
unequal. In that case, the rotation axis for continuous time evolution
is not the bisector of the initial and the target states, even though
it remains in the $\sigma_1-\sigma_3$ plane. As a result, one cannot
reach the target state exactly at any time. The database search succeeds
only with probability less than one, although the rotation angle on the
Bloch sphere for the closest approach to the target state remains $\pi$.
The equal coefficient case is therefore the choice that maximises the
database search success probability.

On the other hand, the results obtained using discrete time evolution for
the Hamiltonian simulation problem are easily extended to the situation
where the two projection operators have unequal coefficients. Without
loss of generality, we can choose
\begin{eqnarray}
\label{uneqmag}
H &=& a|s\rangle\langle s| + |t\rangle\langle t| \\
  &=& \Big(\frac{1+a}{2}\Big)I + \frac{a\sqrt{N-1}}{N}\sigma_1 +
    \Big(\frac{1-a}{2} + \frac{a}{N}\Big)\sigma_3 ~, \nonumber
\end{eqnarray}
with real $a\in[-1,1]$. It gives rise to the evolution operator
(without the global phase)
\begin{equation}
U(t) = \pmatrix{
       \cos(At) - \frac{i}{A}\big(\frac{1-a}{2}+\frac{a}{N}\big)\sin(At)
     & - \frac{i}{A}\big( \frac{a\sqrt{N-1}}{N} \big)\sin(At)            \cr
       - \frac{i}{A}\big( \frac{a\sqrt{N-1}}{N} \big)\sin(At)
     & \cos(At) + \frac{i}{A}\big(\frac{1-a}{2}+\frac{a}{N}\big)\sin(At) \cr
       } ~,
\end{equation}
where
\begin{eqnarray}
A^2 &=& \Big(\frac{1-a}{2}\Big)^2 + \frac{a}{N} \\
    &=& \frac{1}{4}\Big(1-\frac{N-2}{N}a\Big)^2 + \frac{(N-1)a^2}{N^2}
    \geq \Big(\frac{a\sqrt{N-1}}{N}\Big)^2 ~. \nonumber
\end{eqnarray}

The rotation axis for this evolution is still in the $\sigma_1$-$\sigma_3$
plane, and the commutator of the two terms in the Hamiltonian is still
proportional to $H_G$. As a consequence, $U(t)$ can still be expressed
as $Q$ iterations of the Grover operator $U_G$, preceded and followed by
phase rotations:
\begin{equation}
U(t) = \exp\big(i\beta\sigma_3\big) ~(U_G)^Q~
       \exp\Big(i\big(\frac{\pi}{2}+\beta\big)\sigma_3\Big) ~,
\end{equation}
with the parameters given by
\begin{eqnarray}
Q &=& \sin^{-1}\Big( \frac{a\sqrt{N-1}}{AN} \sin(At) \Big)
    \Big/ \big(2\sin^{-1}\frac{1}{\sqrt{N}}\big) ~, \nonumber\\
\beta &=& -\frac{\pi}{4} -\frac{1}{2}
    \tan^{-1}\Big(\big(\frac{1-a}{2}+\frac{a}{N}\big)\frac{1}{A}\tan(At)\Big) ~.
\end{eqnarray}
Note that for $Q<0$, we need to iterate the operator
$U_G^{-1} = - (1-2|t\rangle\langle t|) (1-2|s\rangle\langle s|)$.

\subsection{Discretised Hamiltonian Evolution Complexity}

In a digital implementation, all continuous variables are discretised.
That allows fault-tolerant computation with control over bounded errors.
But it also introduces discretisation errors that must be kept within
specified tolerance level by suitable choices of discretisation intervals.
When Hamiltonian evolution is discretised in time using the Lie-Trotter
formula, the algorithmic error depends on $\Delta t$, which has to be
chosen so as to satisfy the total error bound $\epsilon$ on $U(t)$.
The overall computational complexity is then expressed as a function
of $t$ and $\epsilon$.

For the simplest discretisation, 
\begin{eqnarray}
\label{discreteLieTrot}
\exp\Big(-i\sum_{i=1}^l H_i \Delta t\Big)
&=& \exp\big(-iH_1 \Delta t\big) \ldots \exp\big(-iH_l \Delta t\big) \\
&\times& \exp\big(-iE^{(2)} (\Delta t)^2 \big) , \nonumber
\end{eqnarray}
\begin{equation}
E^{(2)} = \frac{i}{2} \sum_{i<j} [H_i,H_j] + O(\Delta t) ~.
\end{equation}
For the symmetric discretisation,
\begin{eqnarray} 
\exp\Big(-i\sum_{i=1}^l H_i \Delta t\Big)
&=& \Big(\exp(-iH_l \Delta t/2) \ldots \exp(-iH_1 \Delta t/2)\Big) \nonumber\\
&\times& \Big(\exp(-iH_1 \Delta t/2) \ldots \exp(-iH_l \Delta t/2)\Big) \\
&\times& \exp\big(-iE^{(3)} (\Delta t)^3\big) ~, \nonumber
\end{eqnarray}
\begin{eqnarray}
E^{(3)} &=& \frac{1}{24} \sum_{i<j} \big(2[H_i,[H_i,H_j]] + [H_j,[H_i,H_j]]\big)
        \nonumber\\
        &+& \frac{1}{12} \sum_{i<j<k}
            \big(2[H_i,[H_j,H_k]] + [H_j,[H_i,H_k]]\big) + O(\Delta t) ~.
\end{eqnarray}
Here $E^{(k)}$ quantify the size of the discretisation error.
These discretisations maintain exact unitarity, but do not preserve the
energy when $H$ and $E^{(k)}$ do not commute.

For any unitary operator $X$, the norm $||X||$ is equal to one (measured
using either $Tr(X^\dagger X)$ or the magnitude of the largest eigenvalue).
That makes, using Cauchy-Schwarz and triangle inequalities,
\begin{eqnarray}
\label{multistepbound}
||X^m - Y^m|| &=& ||(X-Y)(X^{m-1}+\ldots+Y^{m-1})|| \nonumber\\
              &\le& m||X-Y|| ~.
\end{eqnarray}
So for the total evolution to remain within the error bound $\epsilon_1$, we need
\begin{eqnarray}
\label{multisteperr}
m ||\exp(-iE^{(k)} (\Delta t)^k) - I|| &\approx& m ||E^{(k)}|| (\Delta t)^k \\
  &=& m^{1-k} t^k ||E^{(k)}|| < \epsilon_1 ~. \nonumber
\end{eqnarray}

The error probability can be rapidly reduced by repeating the evolution a
multiple number of times, and then selecting the final result by the majority
rule (not as average). This simple procedure produces an error bound similar
to higher order discretisation formulae. With $R$ repetitions, the error
probability becomes less than $2^{R-1} \epsilon_1^{\lceil R/2 \rceil}$,
which can be made smaller than any prescribed error bound $\epsilon$.%
\footnote{Verification of the result is easy for the database search problem,
          and $R$ repetitions of the algorithm can reduce the error
          probability to less than $\epsilon_1^{-R}$. But verification may
          not be available for the Hamiltonian evolution problem, and so we
          have opted for the majority rule. Majority rule can be applied only
          when the results are discrete. It may be therefore practical to
          postpone the majority voting for the Hamiltonian evolution problem
          to the stage of final determination of the expectation values,
          where the operators $O_a$ can be chosen to have discrete spectra.}
With exact exponentiation of the individual terms $H_i$, the computational
cost to evolve for a single time step $\Delta t$, i.e.  $\cal C$, does not
depend on $\Delta t$. Thus the complexity of the Hamiltonian evolution becomes
\begin{equation}
\label{LieTrotCost}
O(m R {\cal C}) = O\Big(t^{k/(k-1)} \frac{||E^{(k)}||^{1/(k-1)}}
  {\epsilon^{1/((k-1)\lceil R/2\rceil)}} R {\cal C}\Big) ~.
\end{equation}
With superlinear scaling in $t$ and power-law scaling in $\epsilon$, this
scheme based on small $\Delta t$ is not efficient. Note that for the
Hamiltonian $H_C$, $l=2$ is fixed, and both $||E^{(2)}||$ and $||E^{(3)}||$
are $O(N^{-1/2})$. So for evolution time $T=\Theta(N^{1/2})$, the time
complexity becomes linear,
$O(T \epsilon^{-1/((k-1)\lceil R/2\rceil)} R{\cal C})$,
while power-law scaling in $\epsilon$ remains unchanged.

Grover's optimal algorithm uses a discretisation formula where
$\exp(-iH_i \Delta t)$ are reflection operators. The corresponding
time step is large, i.e. $\Delta t_G = \pi$ for Eq.(\ref{LieTrotter})
applied to Eq.(\ref{HamFG}). The large time step introduces another
error because one may jump across the target state during evolution
instead of reaching it exactly. $Q_t$ is not an integer as defined
in Eq.(\ref{evolfracparam}), and needs to be replaced by its nearest
integer approximation $\lfloor Q_t+\frac{1}{2}\rfloor$ in practice.
For instance, the number of time steps needed to reach the target
state in the database search problem is
\begin{equation}
Q = \Big\lfloor \frac{\pi}{2\alpha} \Big\rfloor \approx \frac{\pi}{4}\sqrt{N} ~.
\end{equation}
Since each time step provides a rotation by angle
$\alpha=2\sin^{-1}(1/\sqrt{N})$ along the geodesic in the Hilbert space,
and one may miss the target state by at most half a rotation step, the
error probability of Grover's algorithm is bounded by $\sin^2(\alpha/2)=1/N$.
Since the preceding and following phase rotations in Eq.(\ref{equivevolfrac})
are unitary operations, this error bound applies to $U_C(t)$ as well.
Once again, reducing the error probability with $R$ repetitions of the
evolution and the majority rule selection, we need
$2^{R-1}/N^{\lceil R/2 \rceil} < \epsilon$.
% For $p<1/2$, $\sum_{k=R/2}^R {R \choose k} p^k (1-p)^{R-k}
%              < \sum_{k=R/2}^R {R \choose k} p^{R/2} = 2^{R-1} p^{R/2}$.
% More accurately: R=1,2,3,4,5,6,... give bounds (dropping higher orders)
%                  error < p,2p,3p^2,6p^2,10p^3,20p^3,...
The computational complexity of the evolution is thus
\begin{equation}
\label{Gcomplex}
O(Q_t R {\cal C}_G) = O\Big(\frac{t}{2}
                       \Big(-\frac{2\log\epsilon}{\log N}\Big) {\cal C}_G\Big)
                    = O\Big(-t\frac{\log\epsilon}{\log N} {\cal C}_G\Big) ~.
\end{equation}
With linear scaling in time and logarithmic scaling in $\epsilon$, 
this algorithm is efficient.

It is easy to see why the two algorithms scale rather differently as a
function of $\epsilon$. The straightforward application of the Lie-Trotter
formula makes the time step $\Delta t$ depend on $\epsilon$ as a power-law.
The total error of the algorithm is proportional to the total number of time
steps $m$, and the resultant computational complexity then has a power-law
dependence on $\epsilon$. The power can be reduced by higher order
discretisations or by multiple evolutionary runs and the majority rule
selection, but it cannot be eliminated. On the other hand, with a large time
step that does not depend on $\epsilon$, Grover's algorithm has an error that
is independent of the evolution time. This error is easily suppressed by
multiple evolutionary runs and the majority rule selection. The overall
computational complexity is proportional to the number of evolutionary
runs, which depends only logarithmically on $\epsilon$.

\subsection{Digital State Implementation}
\label{secdigital}

To estimate the computational cost ${\cal C}$, we need to specify quantum
implementation of linear algebra operations involving the block-diagonal
operators $H_i$. It is routine to represent a quantum state in an
$N$-dimensional Hilbert space as
\begin{equation}
|x\rangle = \sum_{j=0}^{N-1} x_j |j\rangle ~,~~
\sum_{j=0}^{N-1} |x_j|^2 = 1 ~,
\end{equation}
where $x_j$ are continuous complex variables. This analog representation
is not convenient for high precision calculations, and so we use the
digital representation instead,%
\footnote{For the same reason, classical digital computers have replaced
          analog computers. It is not possible to measure a physical
          property, say voltage in a circuit, to a million bit precision.
          But that is no obstacle to calculation of, say $\pi$, to a
          million bit precision using digital logic.}
specified by the map
\begin{equation}
\label{digitalrep}
|x\rangle \rightarrow \frac{1}{\sqrt{N}}
\sum_{j=0}^{N-1} |j\rangle |x_j\rangle_b ~.
\end{equation}
This is a quantum state in a $(2^b N)$-dimensional Hilbert space, where
$|x_j\rangle_b$ are the basis vectors of a $b$-bit register representing
the truncated value of $x_j$ (a complex number $x_j$ can be represented
by a pair of real numbers, and $2^b x_j$ are truncated to integers).
This representation is fully entangled between the component index state
$|j\rangle$ and the register value state $|x_j\rangle_b$, with a unique
non-vanishing $|x_j\rangle_b$ (out of $2^b$ possibilities) for every
$|j\rangle$. It is important to observe that no constraint is necessary
on the register values in this representation---the perfect entanglement
ensures unitary evolution in the $(2^b N)$-dimensional space. This freedom
allows simple implementation of linear algebra operations on $|x_j\rangle_b$,
transforming them among the $2^b$ basis states using only C-not and Toffoli
gates of classical reversible logic, with the index state $|j\rangle$ acting
as control. For example,
\begin{eqnarray}
c|x\rangle &\rightarrow&
\frac{1}{\sqrt{N}} \sum_{j=0}^{N-1} |j\rangle |cx_j\rangle_b ~, \\
|x\rangle + |y\rangle &\rightarrow&
\frac{1}{\sqrt{N}} \sum_{j=0}^{N-1} |j\rangle |x_j+y_j\rangle_b ~,
\end{eqnarray}
map non-unitary operations on the left to unitary operations on the right.
The circuits described later in Section \ref{reflser} and Section
\ref{chebyser} combine such elementary operations to construct power series.
Note that a crucial requirement for implementing linear algebra operations
in the digital representation is that only a single index (``$j$" in the
preceding formulae) controls the whole entangled state.

The freedom to choose a convenient representation for the quantum states
is particularly useful due to the fact that the quantum states are never
physically observed. All physically observed quantities are the expectation
values of the form in Eq.(\ref{expectobs}). So to complete the digital
representation, we need to construct for every observable $O_a$ in the
$N$-dimensional Hilbert space a related observable $\tilde{O}_a$ in the
$(2^b N)$-dimensional Hilbert space, such that
\begin{equation}
\langle x| O_a |x \rangle
  = \sum_{j,l=0}^{N-1} x_j^* x_l \langle j| O_a |l \rangle
  = \frac{1}{N}\sum_{j,l=0}^{N-1}
      {}_b\langle x_j| \langle j| \tilde{O}_a |l \rangle |x_l \rangle_b ~.
\end{equation}
For this equality to hold, it suffices to construct the operator
$\tilde{O}_a = O_a \otimes O_b$, where the Hermitian operator $O_b$
in the $2^b$-dimensional Hilbert space satisfies
\begin{equation}
\label{digitalop}
\langle x_j | O_b |x_l \rangle = N x_j^* x_l ~.
\end{equation}
$O_b$ can be looked upon as a metric for the digital register space
in the calculation of expectation values. For a single bit $x_j$,
the solution is easily found to be the measurement operator
$O_{b=1} = N \big(\frac{1-\sigma_3}{2}\big)$. More generally, we note that
\begin{equation}
\langle x_j| (1+\sigma_1)^{\otimes b} |x_l \rangle = 1 ~,
\end{equation}
and the place-value operator for a bit string,
\begin{equation}
V = \sum_{k=0}^{b-1} 2^{-k} I^{\otimes k} \otimes
    \Big(\frac{1-\sigma_3}{2}\Big) \otimes I^{\otimes (b-k-1)} ~,
\end{equation}
gives $V|x_j\rangle = x_j|x_j\rangle$. The solution to Eq.(\ref{digitalop}),
therefore, has a bit-wise fully factorised form, independent of the quantum
state and the observable,%
\footnote{As a matter of fact, any function $f(x_j)$ for the state
          $|x_j\rangle$ can be computed using just the machinery of classical
          reversible logic, and overall normalisations can be adjusted at the
          end of the calculation. Also, note that in terms of the uniform
          superposition state $|s\rangle$,
          $(1+\sigma_1)^{\otimes b} = 2^b|s\rangle\langle s|$.}
\begin{equation}
O_b = N V^\dagger (1+\sigma_1)^{\otimes b} V ~.
\end{equation}
The computational complexity of measurement of physical observables in
the digital representation is thus $O(b^2)$ times that in the analog
representation. The advantages of arbitrary precision calculations and
simple linear algebra, however, unambiguously favour the digital
representation over the analog one.

To efficiently incorporate the digital representation in the Hamiltonian
simulation algorithm, methods must be found to not only manipulate the
register values $|x_j\rangle$ efficiently, but also to initialise and
to observe them. At the start of the calculation, we need to assume that
the initial values $x_j(0)$ can be efficiently computed from $j$. Then the
initial state can be created easily using Hadamard and control operations,
for $N=2^n$, as
\begin{eqnarray}
\label{initstate}
|0\rangle |0\rangle_b
&& ~\mathop{\longrightarrow}\limits^{H^{\otimes n}\otimes I}~
\frac{1}{\sqrt{N}} \sum_{j=0}^{N-1} |j\rangle |0\rangle_b \\
&& ~\mathop{\longrightarrow}\limits^{C_x}~
\frac{1}{\sqrt{N}} \sum_{j=0}^{N-1} |j\rangle |x_j(0)\rangle_b ~.
\end{eqnarray}
When $N$ is not a power of 2, some extra work is needed. A simple fix is
to enlarge the $j$-register to the closest power of 2 and initialise the
additional $x_j$ to zero. Thereafter, the linear algebra operations can be
implemented such that the additional $x_j$ remain zero, and the overall
normalisation (i.e. $1/\sqrt{N}$) can be corrected in the final result
as a proportionality constant. At the end of the calculation, we need
to assume that the final state observables, Eq.(\ref{expectobs}), are
efficiently computable from $x_j(T)$. In such a case, the advantage of the
digital representation is, as pointed out earlier, that the index $j$ can be
handled in parallel (classically) or in superposition (quantum mechanically).

Digital computation with finite register size produces round-off errors,
because real values are replaced by integer approximations. To complete
the analysis, we point out the standard cost estimate to control these
errors. With $b$-bit registers, the available precision is $\delta=2^{-b}$.
Using simple-minded counting, elementary bit-level computational resources
required for additions, multiplications and polynomial evaluations are
$O(b)$, $O(b^2)$ and $O(b^3)$ respectively. (Overflow/underflow limit the
degree of the polynomial to be at most $b$.) All efficiently computable
functions can be approximated by accurate polynomials, so the effort needed
to evaluate individual elements of $H_i$ is $O(b^3)$.

The block-diagonal $H_i$ can be exponentiated exactly. (Depending on the
available quantum logic hardware, Euler angle decomposition may be used
to convert rotations about arbitrary axes to rotations about fixed axes.)
With fixed block sizes, exponential of any block of any $H_i$ can therefore
be obtained to $b$-bit precision with $O(b^3)$ effort. The number of blocks
is $O(N)$, and so the classical cost of multiplying exponential of $H_i$
with a state is proportional to $N$. With an efficient labeling scheme for
the blocks, the index $j$ can be broken down into $O(n)$ tensor product
factors (analogous to Eq.(\ref{initstate})), and then quantum superposition
makes the cost of multiplying exponential of $H_i$ with a state proportional
to $n$. Thus the computational cost of an efficiently encoded matrix-vector
product reduces from its classical scaling $O(Nb^3)$ to its quantum scaling
$O(nb^3)$.

For the database search problem, the number of exponentiations of $H_i$
needed for the Lie-Trotter formula is $m(k-1)l$, which reduces to $2Q_t$
for the Grover version. So with the choice $m(k-1)l\delta=O(\epsilon)$,
i.e. $b=\Omega(\log(m/\epsilon))$, the round-off error can be always made
negligible compared to the discretisation error. The computational cost of
a single evolution step then scales as
\begin{equation}
{\cal C} = O(\log N~(\log(t/\epsilon))^3) = {\cal C}_G ~,
\end{equation}
and the overall evolution complexity of Eq.(\ref{Gcomplex}) becomes
$O(-t~\log\epsilon~(\log(t/\epsilon))^3)$.

\section{From Database Search to More General Projections}
\label{projreflser}

The Hamiltonian $H_C$ for the database search problem is a sum of two
one-dimensional projection operators with equal magnitude. We next construct
an accurate large time step evolution algorithm for the case where the two
projection operators making up the Hamiltonian are more than one-dimensional
but block-diagonal, for example as in Eq.(\ref{oneDlap}). Our strategy now
relies on a rapidly converging series expansion, similar to the proposal
of Ref.~\cite{TalEzer,BCCKS2}, instead of an equivalent Hamiltonian
evolution. We consider, in turn, series expansions in terms of projection
operators and in terms of reflection operators.

\subsection{Algorithm with Projection Operators}

\subsubsection{Series Expansion}

Consider the Hamiltonian decomposition
\begin{equation}
H = H_1 + H_2 ~,~~ H_1^2 = H_1 ~,~~ H_2^2 = H_2 ~.
\end{equation}
Then standard Taylor series expansion around $t=0$ yields
\begin{equation}
\label{twoprojexp}
\exp(-iHt) = I + \sum_{k=1}^\infty c_k(t)
            [ (H_1 H_2 H_1 \ldots)_k + (H_2 H_1 H_2 \ldots)_k ] ~,
\end{equation}
with $c_k(t=0)=0$. Here $(H_1 H_2 H_1 \ldots)_k$ denotes a product of $k$
alternating factors of $H_i$, starting with $H_1$. All other products of
$H_i$ reduce to the two terms retained on r.h.s. in Eq.(\ref{twoprojexp}),
due to the projection operator nature of $H_i$, and the two have the same
coefficients $c_k(t)$ by symmetry. It is worthwhile to observe that the
structure of Eq.(\ref{twoprojexp}) effectively sums up infinite series of
terms---when truncated to order $p$, the series has $2p+1$ terms, compared
to $2^{p+1}-1$ terms in the corresponding series of Ref.~\cite{BCCKS2}.

Differentiating Eq.(\ref{twoprojexp}), we obtain
\begin{eqnarray}
-i(H_1+H_2) &\times& \Big[ I + \sum_{k=1}^\infty c_k(t)
    \left[ (H_1 H_2 H_1 \ldots)_k + (H_2 H_1 H_2 \ldots)_k \right] \Big]
\nonumber\\
&=& \sum_{k=1}^\infty \frac{dc_k(t)}{dt}
    \left[ (H_1 H_2 H_1 \ldots)_k + (H_2 H_1 H_2 \ldots)_k \right] ~,
\end{eqnarray}
which provides the recurrence relation for the coefficients,
\begin{equation}
\frac{dc_k(t)}{dt} = -i(c_k(t) + c_{k-1}(t)) ~,
\end{equation}
with the initial condition $c_0 = 1$. Iterative solution gives
\begin{eqnarray}
\label{coeffval}
c_k(t) &=& (-i)^k e^{-it}
         \int_0^t dt_k \ldots \int_0^{t_3} dt_2 \int_0^{t_2} dt_1~e^{it} 
         \nonumber\\
       &=& (-1)^k e^{-it}
         \left[ e^{it} - \sum_{j=0}^{k-1} \frac{(it)^j}{j!} \right] ~.
\end{eqnarray}
Clearly $|c_k(t)|=O(t^k/k!)$, and we can get as accurate approximations to
$e^{-iHt}$ as desired by truncating Eq.(\ref{twoprojexp}) at sufficiently
high order. Also, the series in Eq.(\ref{twoprojexp}) can be efficiently
summed using nested products, e.g.
\begin{equation}
\sum_{k=1}^p c_k (H_1 H_2 H_1 \ldots)_k
= H_1 (c_1 I + H_2 (c_2 I + H_1 (c_3 I + \ldots))) ~.
\end{equation}

The series in Eq.(\ref{twoprojexp}) can be converted to a form related
to the BCH expansion as:
\begin{equation}
\label{twoprojBCH}
e^{-iHt} = e^{-iH_1 t} e^{-iH_2 t} \Big[ I + \sum_{k=2}^\infty
     \left( c_k^{(1)}(t) (H_1 H_2 H_1 \ldots)_k 
          + c_k^{(2)}(t) (H_2 H_1 H_2 \ldots)_k \right) \Big] ~.
\end{equation}
Noting that $\exp(iH_i t) = I + (e^{it}-1)H_i$, we can evaluate the
coefficients $c_k^{(i)}$ as:
\begin{eqnarray}
c_k^{(1)} &=& e^{it}(c_{k-1} + c_k) - c_{k-1} ~, \cr
c_k^{(2)} &=& (e^{it}-1)^2 (c_{k-2} + c_{k-1}) + c_k^{(1)} ~.
\end{eqnarray}
Although the series in Eq.(\ref{twoprojBCH}) starts with $k=2$,
$c_k^{(i)}(t)$ do not converge any faster than $c_k(t)$ for larger $k$,
and so there is no particular advantage in using it compared to the
series in Eq.(\ref{twoprojexp}).%
\footnote{Choosing the factors on r.h.s. of Eq.(\ref{twoprojBCH}) as
          $e^{-iH_1 t_1} e^{-iH_2 t_2}$, with $t_1=t+O(t^2), t_2=t+O(t^2)$
          allowing some simplification of the series, also does not improve
          the convergence rate of the series.}
% The $k=2$ term is $(1-\cos t)[H_1,H_2] - it(t-\sin t)\{H_1,H_2\}$.

\subsubsection{Complexity Analysis and Series Order Determination}

The computational complexity of Hamiltonian evolution using the Lie-Trotter
formula is $O(m{\cal C})$, as in Eq.(\ref{LieTrotCost}), where $m=t/\Delta t$
and ${\cal C}$ represents the computational cost to evolve for a single time
step. In order to keep the total evolution within the error bound $\epsilon$,
$\Delta t$ has to scale as a power of $\epsilon$, which in turn makes the
computational complexity inefficiently scale as a power of $\epsilon$.
Instead, with a truncated series expansion of Eq.(\ref{twoprojexp}),
we can choose $\Delta t=\Theta(1)$. The order of series truncation, $p$,
is then determined by the error bound $\epsilon$. A single time step needs
$2p$ nested linear algebra operations, with each operation consisting of
a sparse matrix-vector product involving $H_i$, multiplication of a vector
by a constant and addition of two vectors. So the computational complexity
is $O(2mp{\cal C})$ with ${\cal C}$ denoting the computational cost of
evaluating $H_i$ and performing the linear algebra operation. A simple
quantum logic circuit to implement the linear algebra operation, using the
digital representation of Section \ref{secdigital}, is described later in
Section \ref{reflser}.

The series truncation error for a single time step, with $||H_i|| \le 1$
for projection operators, is
\begin{equation}
\Delta[\exp(-iH\Delta t)] \leq 2 \sum_{k=p+1}^\infty |c_k(\Delta t)| ~.
\end{equation}
It has to be bounded by $\epsilon/m$ according to the triangle inequality.
From Eq.(\ref{coeffval}), we have
\begin{equation}
|c_k(\Delta t)| = \left| \sum_{j=k}^\infty \frac{(i\Delta t)^j}{j!} \right|
  \le \frac{(\Delta t)^k}{k!} \left(1 - \frac{\Delta t}{k+1}\right)^{-1} ~.
\end{equation}
The constraint deciding the order of series truncation is, therefore,
\begin{equation}
\label{truncordproj}
2m \frac{(\Delta t)^{p+1}}{(p+1)!}
  \left(1 - \frac{\Delta t}{p+2}\right)^{-2} < \epsilon ~.
\end{equation}
With $\Delta t=\Theta(1)$, we have $m=\Theta(t)$, and the formal solution
is $p=O(\log(t/\epsilon)/\log(\log(t/\epsilon)))$. The computational
complexity of the evolution is then
\begin{equation}
\label{SeriesCost}
O(2mp{\cal C}) = O \left( t \frac{\log(t/\epsilon)}{\log(\log(t/\epsilon))}
                          {\cal C} \right) ~,
\end{equation}
which makes the series expansion algorithm efficient.

Finally, with block-diagonal $H_i$ and finite precision calculations using
$b$-bit registers, the computational cost ${\cal C}$ is $O(nb^3)$.
Then the choice $mp\delta=O(\epsilon)$,
i.e. $b=\Omega(\log(mp/\epsilon))=\Omega(\log((t/\epsilon)\log(t/\epsilon)))$,
makes the round-off errors negligible compared to the truncation error.

\subsubsection{Unequal Magnitude Operators}

The series expansion algorithm is easily extended to the situation where
the two projection operators appearing in the Hamiltonian have unequal
coefficients. With $H=a_1 H_1+a_2 H_2$, the series expansion takes the form
\begin{equation}
\label{twoprojexp1}
\exp(-iHt) = I + \sum_{k=1}^\infty \Big[ c_k(t) (H_1 H_2 H_1 \ldots)_k
               + d_k(t) (H_2 H_1 H_2 \ldots)_k \Big] ~,
\end{equation}
where $c_k(t=0)=0=d_k(t=0)$, and $c_k(t)=d_k(t)$ for even $k$ by symmetry.
Without loss of generality, one may choose $a_1\in[-1,1], a_2=1$ as in
Eq.(\ref{uneqmag}).

Differentiation of Eq.(\ref{twoprojexp1}) leads to the recurrence relations,
\begin{eqnarray}
\frac{dc_k(t)}{dt} &=& -ia_1(c_k(t) + d_{k-1}(t)) ~, \\
\frac{dd_k(t)}{dt} &=& -ia_2(c_{k-1}(t) + d_k(t)) ~,
\end{eqnarray}
with the initial conditions $c_0=1=d_0$. These can be integrated to
\begin{eqnarray}
c_k(t) &=& -ia_1 e^{-ia_1 t} \int_0^t e^{ia_1 t'} d_{k-1}(t') dt' ~, \\
d_k(t) &=& -ia_2 e^{-ia_2 t} \int_0^t e^{ia_2 t'} c_{k-1}(t') dt' ~,
\end{eqnarray}
and iteratively evaluated to any desired order. In particular, for even $k$,
\begin{equation}
c_k(t) = d_k(t) = (a_2 c_{k-1} - a_1 d_{k-1})/(a_1-a_2) ~.
\end{equation}
With rapidly decreasing coefficients, $|c_k(t)|=O(t^k/k!)=|d_k(t)|$,
accurate and efficient truncations of Eq.(\ref{twoprojexp1}) are easily
obtained.

\subsection{Algorithm with Reflection Operators}

\subsubsection{Series Expansion}
\label{reflser}

The series expansion can also be carried out in terms of the reflection
operators $R_i=I-2H_i$, instead of the projection operators $H_i$.
We then have
\begin{eqnarray}
\label{tworeflexp}
e^{it} \exp(-iHt) &=& \exp\Big( i(R_1+R_2) \frac{t}{2} \Big) \nonumber\\
                  &=& r_0(t)~I + \sum_{k=1}^\infty r_k(t)
       \left[ (R_1 R_2 R_1 \ldots)_k + (R_2 R_1 R_2 \ldots)_k \right] ~,
\end{eqnarray}
with $r_0(t=0)=1, r_k(t=0)=0$. The structure of the terms in this series,
with alternating reflection operators, is reminiscent of Grover's algorithm.
Differentiating this expansion, we obtain
\begin{eqnarray}
\frac{i}{2} (R_1+R_2) &\times&
   \Big[ r_0(t)~I + \sum_{k=1}^\infty r_k(t)
   \left[ (R_1 R_2 R_1 \ldots)_k + (R_2 R_1 R_2 \ldots)_k \right] \Big]
   \nonumber\\
&=& \frac{dr_0(t)}{dt} I + \sum_{k=1}^\infty \frac{dr_k(t)}{dt}
    \left[ (R_1 R_2 R_1 \ldots)_k + (R_2 R_1 R_2 \ldots)_k \right] ~,
\end{eqnarray}
which provides the recurrence relations for the coefficients,
\begin{equation}
\label{tworeflrecur}
\frac{dr_0(t)}{dt} = ir_1(t) ~,~~
\frac{dr_k(t)}{dt} = \frac{i}{2}\Big( r_{k+1}(t) + r_{k-1}(t) \Big) ~,
\end{equation}
for $k\geq1$. These are the recurrence relations for the Bessel functions.

Explicit evaluation for the coefficient of identity in the series gives,
using $R_i^2=I$,
\begin{equation}
r_0(t) = \sum_{j=0}^\infty \frac{1}{(2j)!} \Big(\frac{it}{2}\Big)^{2j}
         {2j \choose j} = J_0(t) ~.
\end{equation}
Thereafter, the recurrence relations determine $r_k(t) = i^k J_k(t)$.
With $|r_k(t)|=O(t^k/(2^k k!))$, Eq.(\ref{tworeflexp}) converges
significantly faster than Eq.(\ref{twoprojexp}). It can also be summed
efficiently using nested products. Furthermore, reflections are unitary
operators, and so they are easier to implement in quantum circuits than
projection operators. These properties make Eq.(\ref{tworeflexp}) better
to use in practice than Eq.(\ref{twoprojexp}).

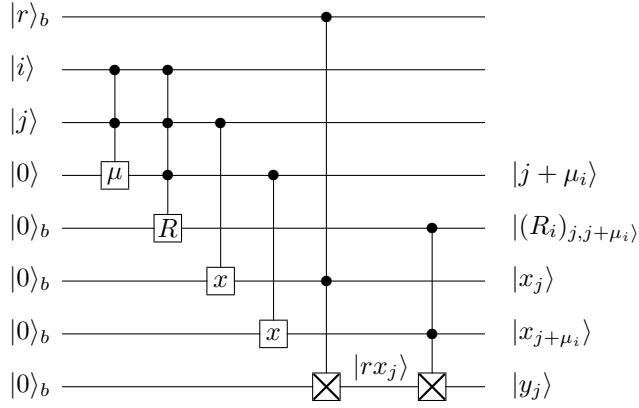
\begin{figure}[t]
{
\begin{center}
%\setlength{\unitlength}{1mm}
% Unit length is 1pt
\begin{picture}(250,140)
\put(20,140){\line(1,0){160}}
\put(20,120){\line(1,0){160}}
\put(20,100){\line(1,0){160}}
\put(20,80){\line(1,0){15}}
\put(45,80){\line(1,0){135}}
\put(20,60){\line(1,0){35}}
\put(65,60){\line(1,0){115}}
\put(20,40){\line(1,0){55}}
\put(85,40){\line(1,0){95}}
\put(20,20){\line(1,0){75}}
\put(105,20){\line(1,0){75}}
\put(20,0){\line(1,0){95}}
\put(125,0){\line(1,0){30}}
\put(165,0){\line(1,0){15}}

\put(0,138){$|r\rangle_b$}
\put(0,118){$|i\rangle$}
\put(0,98){$|j\rangle$}
\put(0,78){$|0\rangle$}
\put(0,58){$|0\rangle_b$}
\put(0,38){$|0\rangle_b$}
\put(0,18){$|0\rangle_b$}
\put(0,-2){$|0\rangle_b$}

\put(40,120){\circle*{4}}
\put(40,100){\circle*{4}}
\put(40,85){\line(0,1){35}}
\put(35,75){\framebox(10,10){$\mu$}}

\put(60,120){\circle*{4}}
\put(60,100){\circle*{4}}
\put(60,80){\circle*{4}}
\put(60,65){\line(0,1){55}}
\put(55,55){\framebox(10,10){$R$}}

\put(80,100){\circle*{4}}
\put(80,45){\line(0,1){55}}
\put(75,35){\framebox(10,10){$x$}}

\put(100,80){\circle*{4}}
\put(100,25){\line(0,1){55}}
\put(95,15){\framebox(10,10){$x$}}

\put(120,140){\circle*{4}}
\put(120,40){\circle*{4}}
\put(120,5){\line(0,1){135}}
\put(115,-5){\framebox(10,10){\huge$\times$}}

\put(160,60){\circle*{4}}
\put(160,20){\circle*{4}}
\put(160,5){\line(0,1){55}}
\put(155,-5){\framebox(10,10){\huge$\times$}}

\put(130,5){$|rx_j\rangle$}
\put(190,78){$|j+\mu_i\rangle$}
\put(190,58){$|(R_i)_{j,j+\mu_i\rangle}$}
\put(190,38){$|x_j\rangle$}
\put(190,18){$|x_{j+\mu_i}\rangle$}
\put(190,-2){$|y_j\rangle$}

\end{picture}
\end{center}
}
\caption{Digital quantum logic circuit for the linear algebra fragment
$|y\rangle = (rI + R_i)|x\rangle$ occurring in the nested evaluation of
the series in Eq.(\ref{tworeflexp}). Among the controlled logic gates,
\frame{\strut$~\mu~$}, \frame{\strut$~R~$} and \frame{\strut$~x~$} denote
oracle operations specified by the Hamiltonian and the initial state,
while \frame{\large$\times$} stands for the generalised Toffoli gate
implementing $|a,b,c\rangle \rightarrow |a,b,c+ab\rangle$. The circuit
is used with a uniform superposition over the index $j$.}
\label{fragseries}
\end{figure}

Summation of the series in Eq.(\ref{tworeflexp}), truncated to order $p$,
by nested products requires $2p$ executions of the elementary linear
algebra operation fragment $(rI+R)|x\rangle$. Each fragment contains
three simple components: multiplication of a vector by a unitary matrix,
multiplication of a vector by a constant, and addition of two vectors.
Its evaluation using the digital representation of Section \ref{secdigital}
is schematically illustrated in Fig.\ref{fragseries}. Multiplication of
a vector by a diagonal matrix, and addition of two vectors are easy tasks.
Multiplication of $|x\rangle$ by the off-diagonal elements of $R_i$ needs
a little care, and can be accomplished by shuffling the elements of
$|x\rangle$. Since $R_i$ are block-diagonal, this shuffling is only within
each block, and requires a fixed number of permutations that depend on the
block size but not on the system size. The time complexity of the series
summation is thus $O(p{\cal C})$, where ${\cal C}=O(nb^3)$ using quantum
superposition over the index $j$. The space resources required to combine
together the results of all the fragments are a fixed number of $n$-bit
registers and $O(p)$ $b$-bit registers. (The registers used for off-diagonal
matrix multiplication in individual fragments can be reversibly restored to
zero, and then reused in subsequent steps.) These features make the algorithm
efficient, and the procedure is considerably simpler than the corresponding
series summation method in Ref.~\cite{BCCKS2}.

\subsubsection{Complexity Analysis and Series Order Determination}
\label{reflord}

When the series of Eq.(\ref{tworeflexp}) is truncated at order $p$,
with time step $\Delta t$ and $||R_i||=1$, the truncation error is
\begin{equation}
\Delta[\exp(-iH\Delta t)] \leq 2 \sum_{k=p+1}^\infty |r_k(\Delta t)| ~.
\end{equation}
Since the Bessel functions obey
\begin{equation}
\label{Bessfn}
J_k(z) = \sum_{s=0}^\infty \frac{(-1)^s (z/2)^{k+2s}}{s! (k+s)!}
       = \frac{z^k}{2^k k!} \left( 1+O\Big(\frac{z^2}{k}\Big) \right) ~,
\end{equation}
it follows that (assuming $(\Delta t)^2 \leq 8(p+2)$) :%
\footnote{For an alternating series, with successive terms decreasing
          monotonically in magnitude, the leading omitted term provides
          a bound on the truncation error.}
\begin{equation}
\sum_{k=p+1}^\infty |J_k(\Delta t)|
  \leq \sum_{k=p+1}^\infty \frac{(\Delta t)^k}{2^k k!}
%      \exp\Big( \frac{(\Delta t)^2}{4(k+1)} \Big)
  \leq \frac{(\Delta t)^{p+1}}{2^{p+1}(p+1)!}
%      \exp\Big( \frac{(\Delta t)^2}{4(p+2)} \Big)
       \left( 1-\frac{\Delta t}{2(p+2)} \right)^{-1} .
\end{equation}
With $t=m\Delta t$, the order of series truncation is therefore decided
by the constraint
\begin{equation}
\label{truncordrefl}
2m\frac{(\Delta t)^{p+1}}{2^{p+1}(p+1)!}
% \exp\Big( \frac{(\Delta t)^2}{4(p+2)} \Big)
  \left( 1-\frac{\Delta t}{2(p+2)} \right)^{-1} < \epsilon ~.
\end{equation}
For time step $\Delta t=\Theta(1)$, the formal solution is again
$p=O(\log(t/\epsilon)/\log(\log(t/\epsilon)))$. That keeps the algorithm
efficient, with the same computational complexity as in Eq.(\ref{SeriesCost}).

\subsubsection{Unequal Magnitude Operators}

When the two reflection operators have unequal coefficients in the
Hamiltonian, we can expand
\begin{eqnarray}
\label{tworeflexp1}
\exp\Big( i(a_1 R_1+a_2 R_2) \frac{t}{2} \Big) &=& p_0(t)~I \\
&+& \sum_{k=1}^\infty \left[ p_k(t) (R_1 R_2 R_1 \ldots)_k
    + q_k(t) (R_2 R_1 R_2 \ldots)_k \right] ~, \nonumber
\end{eqnarray}
with $p_0(t=0)=1, p_k(t=0)=0=q_k(t=0)$, and $p_k(t)=q_k(t)$ for even $k$
by symmetry.

Differentiation of Eq.(\ref{tworeflexp1}) leads to the recurrence relations,
\begin{eqnarray}
\frac{dp_k(t)}{dt} &=& \frac{i}{2}\Big( a_1 q_{k-1}(t) + a_2 q_{k+1}(t) \Big) ~, \\
\frac{dq_k(t)}{dt} &=& \frac{i}{2}\Big( a_2 p_{k-1}(t) + a_1 p_{k+1}(t) \Big) ~,
\end{eqnarray}
for $k\geq1$. These can be iteratively solved to obtain the coefficients
$p_k$ and $q_k$ for $k\geq2$, to any desired accuracy, starting from the
initial coefficients $p_0=q_0$, $p_1$ and $q_1$. Explicit evaluation of
these initial coefficients gives
\begin{equation}
p_0(t) = \sum_{j=0}^\infty \frac{1}{(2j)!} \Big(\frac{it}{2}\Big)^{2j}
         \left( \sum_{l=0}^j {j \choose l}^2 a_1^{2(j-l)} a_2^{2l} \right) ~,
\end{equation}
\begin{equation}
p_1(t) = \sum_{j=0}^\infty \frac{1}{(2j+1)!} \Big(\frac{it}{2}\Big)^{2j+1}
       \times  \left( \sum_{l=0}^j {j \choose l} {j+1 \choose l}
                             a_1^{2(j-l)+1} a_2^{2l} \right) ~.
\end{equation}
$q_1(t)$ is obtained from $p_1(t)$ by interchanging $a_1 \leftrightarrow a_2$,
and we also have the relation:
\begin{equation}
\frac{dp_0(t)}{dt} = \frac{i}{2}\Big( a_1 p_1(t) + a_2 q_1(t) \Big) ~.
\end{equation}
The bounds $|p_k(t)|=O(t^k/(2^k k!))=|q_k(t)|$ make accurate and efficient
truncations of Eq.(\ref{tworeflexp1}) possible.

\subsection{Numerical Tests}
\label{numtest}

The computational complexity bounds, Eq.(\ref{LieTrotCost}) and 
Eq.(\ref{SeriesCost}), have been obtained assuming that the evolution
errors during different time steps are unrelated. In practice, these
bounds are not tight because correlations exist between evolution errors
at successive time steps. To judge the tightness of the bounds, and also
to estimate the scaling coefficients involved, we simulated the Lie-Trotter
(with $k=2$) and series expansion algorithms, Eqs.(\ref{discreteLieTrot})
and (\ref{twoprojexp},\ref{tworeflexp}) respectively, for the one-dimensional
discretised Laplacian $(H_e+H_o)/2$ defined as per Eq.(\ref{oneDlap}).

\begin{figure}[b!]
\begin{center}
\includegraphics[width=9cm]{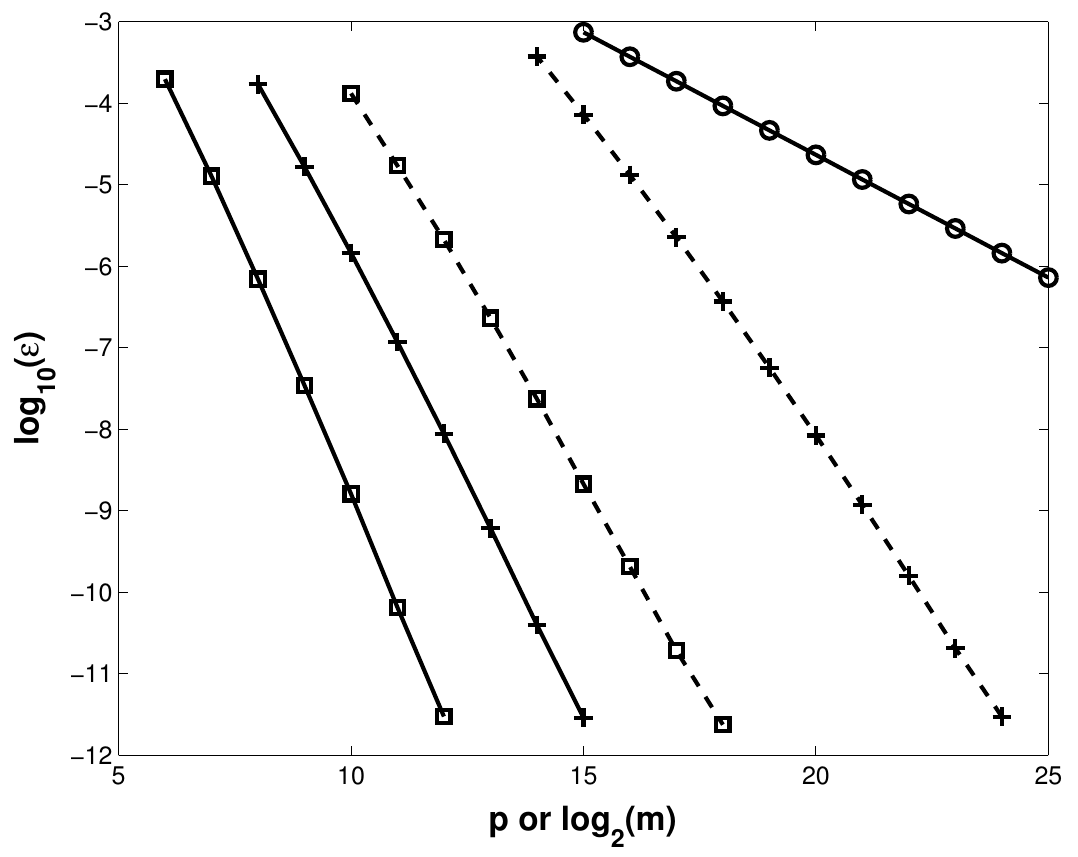}
\end{center}
\caption{Dependence of the error $\epsilon$ on the truncation order $p$
for the series expansion algorithms, and $\log_2 m$ for the Lie-Trotter
algorithm. The symbols $\boxdot$, $+$ and $\odot$ respectively represent
the results for the reflection operator series, the projection operator
series and the Lie-Trotter algorithm. Continuous and dashed lines connect
series expansion results for $\Delta t=1$ and $\Delta t=\pi$ respectively.}
\label{errorscaling}
\end{figure}

We carried out our tests on a one-dimensional periodic lattice of length
$L=128$, with a random initial state $|\psi(0)\rangle$. We quantified the
error as the norm of the difference between the simulated and the exact
states, i.e. $\epsilon=\big|\big| |\widetilde{\psi}\rangle-|\psi\rangle
\big|\big|$. We also needed $mp\delta < \epsilon$ to keep the round-off
errors under control. That was not possible with 32-bit arithmetic, and
we used 64-bit arithmetic.

\begin{figure}[t]
\begin{center}
\includegraphics[width=9cm]{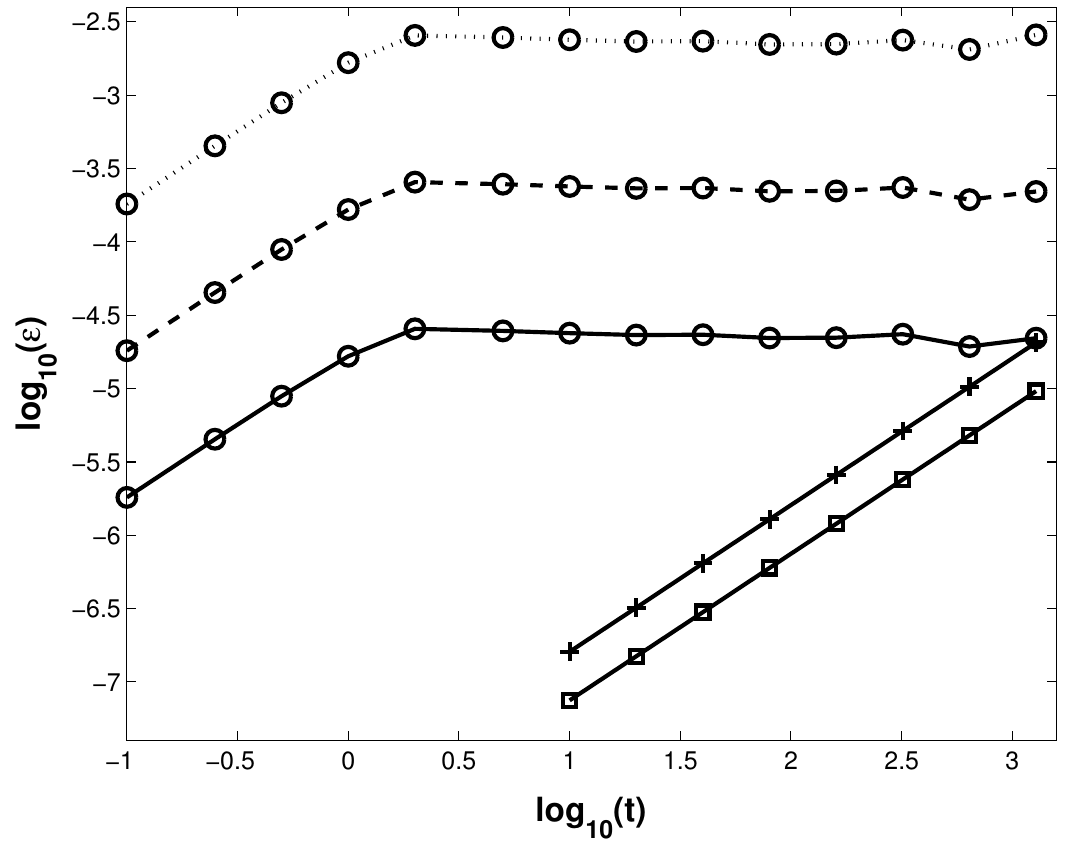}
\end{center}
\caption{Dependence of the error $\epsilon$ on the evolution time for the
series expansion and the Lie-Trotter algorithms. The symbols $\boxdot$,
$+$ and $\odot$ respectively represent the results for the reflection
operator series (with $p=8,\Delta t=1)$, the projection operator series
(with $p=10, \Delta t=1)$, and the Lie-Trotter algorithm (with
$\Delta t=0.0001$, $0.001$ and $.01$ values connected by continuous,
dashed and dotted lines respectively).}
\label{timescaling}
\end{figure}

We selected $t=100$ to study the dependence of the error on the evolution
step size and the series truncation order. Our results are displayed in
Fig.\ref{errorscaling}. For the series expansion algorithms, as expected,
we observe that (a) the truncation order $p$ depends linearly on
$\log(\epsilon)$, (b) the numerical values are consistent with the bounds
in Eqs.(\ref{truncordproj},\ref{truncordrefl}) but the bounds are not very
tight, and (c) the reflection operator series converges faster than the
projection operator series. For $\Delta t=1$, the coefficients
$c_k(\Delta t)$ and $r_k(\Delta t)$ decrease monotonically, and the series
reach a given error $\epsilon$ with a smaller order $p$ compared to the
case $\Delta t=\pi$. But in the overall computational complexity, this
reduction in $p$ (roughly a factor of 1.6) is more than offset by the
increase in $m$ (a factor of $\pi$), and so the choice $\Delta t=\pi$ is
slightly more efficient (by roughly a factor of 2). Even larger $\Delta t$
increase the range over which $c_k(\Delta t),r_k(\Delta t)$ vary, and hence
require higher precision arithmetic (i.e. larger $b$). Consequently, it may
not be practical to implement such large $\Delta t$.

For the Lie-Trotter algorithm, we find that the error $\epsilon$ is
inversely proportional to $m$. As a specific comparison, to make
$\epsilon<10^{-5}$, we needed $p>16$ for the projection operator series,
$p>11$ for the reflection operator series (both with $\Delta t=\pi$),
and $m>2^{21}$ for the Lie-Trotter algorithm. The computational cost
$2mp{\cal C}$ of the series expansion algorithms is then of the order
$7 \times 10^2{\cal C}$, which is a huge improvement over the corresponding
cost $ml{\cal C} = 4\times 10^6{\cal C}$ for the Lie-Trotter algorithm.
The ratio of the two is consistent with the order of magnitude expectation
$(-\epsilon\log\epsilon)$.

To study the growth of the error with the evolution time, we varied the
simulation time $t$, while holding $\Delta t$ and $p$ fixed. Our results
are illustrated by Fig.\ref{timescaling}. For the series expansion
algorithms, we find that $\epsilon$ is proportional to $t$, implying that
the errors of successive time steps additively accumulate, in accordance
with Eq.(\ref{multistepbound}). But we also find that for the Lie-Trotter
algorithm such additive accumulation of error holds only for $t\lesssim1$.
Beyond that the error saturates with the saturation value proportional
to $\Delta t$. This stoppage of error growth for large $t$ indicates
cancellations among the errors of different time steps, possibly due to
correlations in the periodic evolution beyond the first cycle (period
of $\exp(-iH_i t)$ is $2\pi$).%
\footnote{We are unable to figure out whether the error saturation is
          specific to our choice of the evolution Hamiltonian,
          Eq.(\ref{oneDlap}), or whether it would hold for more general
          Hamiltonians as well.}
We note that for $t\gtrsim 1$, we have roughly $m=t/\Delta t\propto t/\epsilon$,
and not $m \propto t^2/\epsilon$ as per Eq.(\ref{multisteperr}).

\section{Efficient Simulation of Local Hamiltonian Evolution}
\label{localham}

We now construct a rapidly converging series expansion for $\exp(-iHt)$,
where $H$ is any local efficiently computable Hamiltonian. (The reason for
decomposing the Hamiltonian into block-diagonal parts appears later in
Section \ref{chebyser}.) It is well-known that an expansion in terms of
the Chebyshev polynomials provides uniform approximation for any bounded
function, with fast convergence of the series \cite{Arfken}. We use such
an expansion for $\exp(-iHt)$, interpreting all matrix functions as their
power series expansions \cite{TalEzer}.

\subsection{Chebyshev Expansion and its Complexity}

For any bounded Hamiltonian, its eigenvalue spectrum is within a range
$[\lambda_{\rm min},\lambda_{\rm max}]$. With a linear transformation,
this range can be mapped to the interval $[-1,1]$ that is the domain
of the Chebyshev polynomials $T_n(x)=\cos(n\cos^{-1}x)$. Explicitly,
\begin{eqnarray}
e^{-iHt} &=& e^{-i(\lambda_{\rm max}+\lambda_{\rm min})t/2}
             e^{-i\tilde{H}\tilde{t}} ~, \\
\tilde{H} &=& (2H - (\lambda_{\rm max}+\lambda_{\rm min})I)
             /(\lambda_{\rm max}-\lambda_{\rm min}) , \\
\tilde{t} &=& t (\lambda_{\rm max}-\lambda_{\rm min})/2 ~.
\label{tscal}
\end{eqnarray}
In situations where $\lambda_{\rm min}$ and $\lambda_{\rm max}$ are not
exactly known, respectively lower and upper bounds for them can be used.
Henceforth, we assume that such a mapping has been carried out and drop
the tilde's on $H$ and $t$ for simplicity.

The Chebyshev expansion gives
\begin{equation}
\label{chebyexp}
e^{-iHt} = \sum_{k=0}^\infty C_k(t)~T_k(H) ~,
\end{equation}
where the expansion coefficients are the Bessel functions:
\begin{eqnarray}
C_0     &=& \frac{1}{\pi} \int_0^\pi e^{-it\cos\theta} d\theta = J_0(t) ~, \\
C_{k>0} &=& \frac{2}{\pi} \int_0^\pi e^{-it\cos\theta} \cos(k\theta) ~d\theta 
         =  2(-i)^k J_k(t) ~.
\end{eqnarray}
Note that the Chebyshev polynomials are bounded in $[-1,1]$, and the
coefficients $J_k(t) = t^k/(2^k k!)+\ldots$ fall off faster by a factor
of $2^k$ compared to the corresponding coefficients $t^k/k!$ of the
Taylor series expansion. This is the well-known advantage of the Chebyshev
expansion compared to other series expansions.

For the special case analysed in Section \ref{reflser}, i.e.
$H=I-(R_1+R_2)/2$, $R_i^2 = I$ implies that the spectrum of $H$ is
bounded in $[-1,1]$. Then from the recursion relation for the Chebyshev
polynomials,
\begin{equation}
T_{k+1}(H) = 2H~T_k(H) - T_{k-1}(H) ~,
\label{ChebyRecur}
\end{equation}
it follows that
\begin{equation}
T_k\left( -\frac{R_1+R_2}{2} \right) = \frac{(-1)^k}{2}
\Big( (R_1 R_2 \ldots)_k + (R_2 R_1 \ldots)_k \Big) .
\end{equation}
Thus we reproduce the series expansion obtained in Eq.(\ref{tworeflexp}),
with the same values for $r_k(t)$. Looking at it another way, \emph{the
partial summation of the reflection operator series for $e^{-iHt}$ converts
the Taylor series expansion into a better behaved Chebyshev expansion}.

When the Chebyshev expansion in Eq.(\ref{chebyexp}) is truncated at order
$p$, its error analysis is identical to that in Section \ref{reflord}.
Choosing $t=m\Delta t$ and $\Delta t=\Theta(1)$, the constraint of 
Eq.(\ref{truncordrefl}) formally provides an efficiently converging
series with $p=O(\log(t/\epsilon)/\log(\log(t/\epsilon)))$.

We have noted earlier that reflection operations are the largest evolution
steps consistent with unitarity, and their use makes Grover's algorithm
optimal. Then, with $e^{i\pi R/2}=iR$, a good guess for the evolution
time step is $\Delta t=\pi$. With this choice, the numerical tests in
Section \ref{numtest} indicate that truncating the series at order
$p=2\ln(t/\epsilon)/\ln(\ln(t/\epsilon))$ is sufficiently accurate.

A truncated series of the Chebyshev expansion is efficiently evaluated using
Clenshaw's algorithm, based on the recursion relation Eq.(\ref{ChebyRecur}).
One initialises the vectors
$|y_{p+1}\rangle = 0$, $|y_p\rangle = C_p|x\rangle$,
and then uses the reverse recursion
\begin{equation}
\label{clenshawrec}
|y_k\rangle = C_k|x\rangle + 2H~|y_{k+1}\rangle - |y_{k+2}\rangle ~,
\end{equation}
from $k=p-1$ to $k=0$. At the end,
\begin{equation}
\sum_{k=0}^p C_k T_k(H) |x\rangle =
(C_0|x\rangle + |y_0\rangle - |y_2\rangle)/2
\end{equation}
is obtained using $p$ sparse matrix-vector products involving $H$.
The computational complexity of the total evolution is then
\begin{equation}
\label{Chebycomplexity}
O(mp{\cal C}_C) = O\left( t\frac{\log(t/\epsilon)}
                                {\log(\log(t/\epsilon))}{\cal C}_C \right) ~,
\end{equation}
where ${\cal C}_C$ is the computational cost of implementing the recursion
of Eq.(\ref{clenshawrec}).

\subsection{An Alternate Strategy}

The Chebyshev expansion coefficients $J_k(t)$ are bounded for any value of
$t$, unlike their Taylor series counterparts, and rapidly fall off for $k>t$.
These properties suggest an alternate evolution algorithm, i.e. evaluate
$e^{-iHt}$ at one shot without subdividing the time interval into multiple
steps \cite{TalEzer}. Of course, this requires the Hamiltonian to be time
independent; otherwise, the evolution has to be performed piece-wise over
time intervals within which the Hamiltonian is effectively constant.

The error due to truncating the Chebyshev expansion at order $p$ is bounded by
\begin{equation}
\label{onestepbound}
\sum_{k=p+1}^\infty |C_k(t)|
  \leq \sum_{k=p+1}^\infty \frac{t^k}{2^{k-1} k!}
  \leq \frac{t^{p+1}}{2^p (p+1)!} \left( 1 - \frac{t}{2(p+2)} \right)^{-1} ,
\end{equation}
provided the subleading contribution in Eq.(\ref{Bessfn}) can be ignored.
The subleading contribution can certainly be neglected for $p>t^2/8$, but
the bound in Eq.(\ref{onestepbound}) may hold for even smaller values of $p$
due to cancellations among subleading contributions of different terms in
the series. Making Eq.(\ref{onestepbound}) smaller than $\epsilon$ requires
$p+1>et/2$, and the formal bound is
$p=O(t\epsilon^{-2/(et)})=O(t+\log(1/\epsilon))$ for $t>\log(1/\epsilon)$.
The resultant computational complexity of the evolution,
\begin{equation}
O(p{\cal C}_C) = O(t \epsilon^{-2/(et)} {\cal C}_C)
               = O((t+\log(1/\epsilon)) {\cal C}_C) ~,
\end{equation}
can be comparable to Eq.(\ref{Chebycomplexity}) for values of $\epsilon$
and $t$ that are of practical interest. The extent to which the computational
complexity would be enhanced by the need to control subleading contributions
can be problem dependent, and needs to be determined numerically \cite{TalEzer}.

To implement this strategy, the Bessel functions $J_k(t)$ upto order $p$
need to be evaluated to $b=\Omega(\log(p/\epsilon))$ bit precision. That
can be efficiently accomplished using the recursion relation,
\begin{equation}
J_{k-1}(t) = \frac{2k}{t} J_k(t) - J_{k+1}(t) ~,
\end{equation}
in descending order \cite{AMS55}. One starts with approximate guesses for
$J_l(t)$ and $J_{l+1}(t)$, with $l$ slightly larger than $p$, and uses the
recursion relation repeatedly to reach $J_0(t)$. Then all the values are
scaled to the correct normalisation by imposing the constraint
$J_0(t) + 2\sum_{k=1}^{\lceil l/2 \rceil} J_{2k}(t) = 1$. This procedure to
determine the expansion coefficients requires $\Theta(pb^2)$ computational
effort, and so does not alter the overall computational complexity.

\subsection{Digital State Implementation}
\label{chebyser}

Summation of the series in Eq.(\ref{chebyexp}), truncated to order
$p$, requires $p$ executions of the Clenshaw recursion relation,
Eq.(\ref{clenshawrec}). Multiplication of a vector by a constant,
and addition of two vectors, are easily carried out with the digital
representation of Section \ref{secdigital}. Multiplication of the
sparse Hamiltonian with a vector, on the other hand, has to be
carefully implemented such that quantum parallelism converts its
computational complexity from classical $O(N)$ to quantum $O(n)$. 

Multiplication by the diagonal elements of the Hamiltonian has a trivial
parallel structure. but its parallelisation for the off-diagonal elements
of the Hamiltonian needs decomposition of $H$ into parts, with each part
consisting of a large number of mutually independent blocks. As mentioned
earlier in Section \ref{hamdecomp}, such a decomposition can be achieved
for any sparse Hamiltonian using an edge-colouring algorithm for the
corresponding graph. With $l$ colours, there are $l$ Hamiltonian parts,
each containing $O(N/2)$ mutually independent $2\times2$ blocks. (Note
that Hermiticity of the Hamiltonian relates the off-diagonal elements,
$H_{j,j+\mu} = H^*_{j+\mu,j}$, that are represented by a single edge of
the graph.) Evaluating the contribution of each Hamiltonian part in
succession, and combining the individual block calculations for each
Hamiltonian part with a superposition of their block labels, the total
computational effort for Hamiltonian multiplication becomes $O(l\log(N/2))$
times the effort for a single $2\times2$ block multiplication.

In the digital representation, the $2\times2$ block multiplication becomes
straightforward provided one can swap the $b$-bit register values, i.e.
\begin{equation}
\label{swapy}
|j\rangle |y_j\rangle + |j+\mu\rangle |y_{j+\mu}\rangle \longrightarrow
|j\rangle |y_{j+\mu}\rangle + |j+\mu\rangle |y_j\rangle ~.
\end{equation}
Such a swap operation can be performed by the reflection operator,
\begin{equation}
S = \sigma_1 \otimes I^{\otimes b} ~,~~ S^2 = I ~,
\end{equation}
acting on the subspace
$\{|j\rangle,|j+\mu\rangle\} \otimes \{|y_j\rangle,|y_{j+\mu}\rangle\}$.
The swap can be easily undone after the off-diagonal element multiplication
for a particular Hamiltonian part $H_i$, to use $|y_j\rangle$ again for the
next Hamiltonian part.

\begin{figure}[t]
{
\begin{center}
%\setlength{\unitlength}{1mm}
% Unit length is 1pt
\begin{picture}(250,180)
\put(20,180){\line(1,0){35}}
\put(65,180){\line(1,0){10}}
\put(85,180){\line(1,0){50}}
\put(145,180){\line(1,0){30}}
\put(185,180){\line(1,0){5}}
\put(25,160){\line(1,0){165}}
\put(25,140){\line(1,0){165}}
\put(20,120){\line(1,0){170}}
\put(20,100){\line(1,0){170}}
\put(20,80){\line(1,0){15}}
\put(45,80){\line(1,0){145}}
\put(20,60){\line(1,0){35}}
\put(65,60){\line(1,0){125}}
\put(20,40){\line(1,0){95}}
\put(125,40){\line(1,0){60}}
\put(40,20){\line(1,0){55}}
\put(105,20){\line(1,0){50}}
\put(165,20){\line(1,0){25}}
\put(40,0){\line(1,0){150}}

\put(0,178){$|0\rangle_b$}
\put(0,158){$|C_k\rangle_b$}
\put(0,138){$|x_j\rangle_b$}
\put(0,118){$|i\rangle$}
\put(0,98){$|j\rangle$}
\put(0,78){$|0\rangle$}
\put(0,58){$|0\rangle_b$}
\put(0,38){$|0\rangle_b$}
\put(0,18){$|(y_{k+1})_j\rangle_b$}
\put(0,-2){$|(y_{k+2})_j\rangle_b$}

\put(40,120){\circle*{4}}
\put(40,100){\circle*{4}}
\put(40,85){\line(0,1){35}}
\put(35,75){\framebox(10,10){$\mu$}}

\put(60,160){\circle*{4}}
\put(60,140){\circle*{4}}
\put(60,140){\line(0,1){35}}
\put(55,175){\framebox(10,10){{\huge$\times$}}}

\put(60,120){\circle*{4}}
\put(60,100){\circle*{4}}
\put(60,65){\line(0,1){55}}
\put(55,55){\framebox(10,10){$H$}}

\put(80,20){\circle*{4}}
\put(80,60){\circle*{4}}
\put(80,20){\line(0,1){155}}
\put(75,175){\framebox(10,10){{\huge$\times$}}}

\put(100,100){\circle*{4}}
\put(100,80){\circle*{4}}
\put(100,25){\line(0,1){75}}
\put(95,15){\framebox(10,10){$S$}}

\put(120,120){\circle*{4}}
\put(120,100){\circle*{4}}
\put(120,80){\circle*{4}}
\put(120,45){\line(0,1){75}}
\put(115,35){\framebox(10,10){$H$}}

\put(140,20){\circle*{4}}
\put(140,40){\circle*{4}}
\put(140,20){\line(0,1){155}}
\put(135,175){\framebox(10,10){{\huge$\times$}}}

\put(160,100){\circle*{4}}
\put(160,80){\circle*{4}}
\put(160,25){\line(0,1){75}}
\put(155,15){\framebox(10,10){$S$}}

\put(180,0){\circle*{4}}
\put(180,0){\line(0,1){175}}
\put(175,175){\framebox(10,10){\large$-$}}

\put(107,10){$|(y_{k+1})_{j+\mu_i}\rangle$}
\put(195,178){$|(y_k)_j\rangle_b$}
\put(195,78){$|j+\mu_i\rangle$}
\put(195,58){$|(2H_i)_{j,j}\rangle$}
\put(190,38){$|(2H_i)_{j,j+\mu_i}\rangle$}
\put(195,18){$|(y_{k+1})_j\rangle$}

\end{picture}
\end{center}
}
\caption{Digital quantum logic circuit for executing the recursion relation
of Clenshaw's algorithm, Eq.(\ref{clenshawrec}), to be executed with a
uniform superposition over the index $j$. Operations for a single $H_i$
containing only $2\times2$ blocks (labeled by $j,j+\mu_i$) are shown.
Among the controlled logic gates, \frame{\strut$~\mu~$} and
\frame{\strut$~H~$} denote oracle operations specified by the Hamiltonian,
\frame{\strut$~S~$} is the swap operation of Eq.(\ref{swapy}), 
\frame{\large$\times$} stands for the generalised Toffoli gate
implementing $|a,b,c\rangle \rightarrow |a,b,c+ab\rangle$, and
\frame{\large$-$} labels the generalised C-not gate performing
$|a,b\rangle \rightarrow |a,b-a\rangle$.}
\label{fragcheby}
\end{figure}
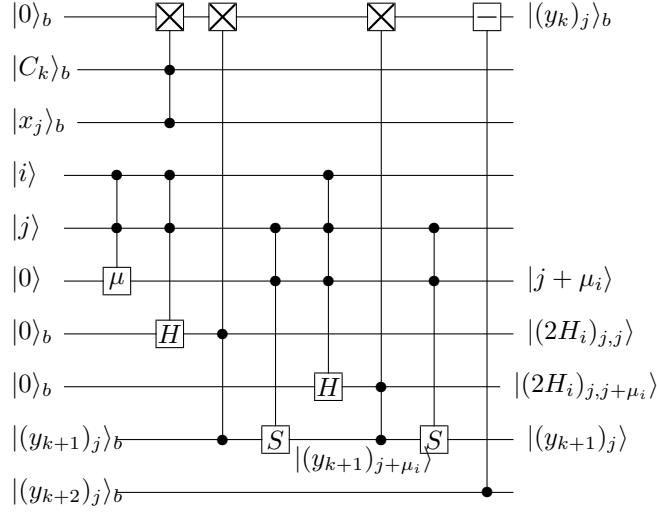

The digital circuit implementation of Eq.(\ref{clenshawrec}), for a single
Hamiltonian part $H_i$, is schematically illustrated in Fig.\ref{fragcheby}.
It has computational complexity $O(b^3)$ arising from evaluation of the
Hamiltonian elements; the rest of the linear algebra operations have
computational complexity $O(b^2)$. Including contributions of all the
Hamiltonian parts, and the computational effort needed to superpose the
index $j$, we thus have the time complexity ${\cal C}_C=O(lnb^3)$.
We also point out that the space resources required to put together the
full Chebyshev expansion are a fixed number of $n$-bit registers and $O(p)$
$b$-bit registers.

Finally, note that the classical computational complexity for implementing
Eq.(\ref{clenshawrec}) is ${\cal C}_C=O(lNb^3)$. \emph{In our construction
based on digital representation for the quantum states, the full quantum
advantage that reduces $N$ to $n$ arises from a simple superposition of
the quantum state label $j$, and this superposition in turn requires
decomposition of the Hamiltonian into block-diagonal parts.}

\section{Summary and Outlook}

We have presented efficient quantum Hamiltonian evolution algorithms
belonging to the class P:P, for local efficiently computable Hamiltonians
that can be mapped to graphs with bounded degree. Our construction exploits
the fact that, the Lie-Trotter evolution formula can be reorganised in terms
of reflection operators and Chebyshev expansions (by partially summing up
the BCH or the Taylor expansions), so as to be accurate for finite time step
size $\Delta t=\Theta(1)$. Specifically, $P^2=P$ and $R^2=I$ allow easy
summation of a large number of terms, while the large spectral gap of $P$
and $R$, due to only two distinct eigenvalues, provides a rapid convergence
of the series.%
\footnote{Among all operators with unit norm, the reflection operators with
          eigenvalues $\pm1$ have the largest spectral gap. They are used
          in Grover's optimal algorithm, and they are our best expansion
          components.}
The net result is a dramatic exponential gain in the computational error
complexity. Our expansions have better convergence properties than previous
similar results \cite{BCCKS1,BCCKS2}, obtained by successively reducing the
Hamiltonian evolution problem to simpler instances. Furthermore, our explicit
constructions show how to design practical efficient algorithms, and reveal
the physical reasons underlying their efficiency.

The formalism that we have developed has connections to the familiar method
for combining exponentials of operators, i.e. the Baker-Campbell-Hausdorff
formula. This formula can be partially summed up and simplified for
exponentials of projection operators. Several identities for projection
operators that are useful in the process are described in the Appendix.
In particular, the identity of Eq.(\ref{adPident}) may be useful in other
applications of the BCH expansion.

Our methods have introduced two concepts that go beyond the specific problem
investigated here. One is that unitary time evolution using a large step size
can be looked upon as simulation of an effective Hamiltonian. This effective
Hamiltonian can be very different from the original Hamiltonian that defined
the evolution problem in continuous time, as seen in our analysis of Grover's
algorithm. Such a correspondence between two distinct Hamiltonians that give
the same finite time evolution is highly non-trivial, and underlies efficient
summation of the BCH expansion. The technique of speeding up simulations by
finding appropriate equivalent Hamiltonians can be useful in a variety of
problems defined as continuous time evolutions (including adiabatic ones).

The other novel concept we have used is to map non-unitary linear algebra
operations to unitary operators using the digital state representation.
High precision calculations need a digital representation instead of an
analog one. We have introduced such a representation for both the quantum
states and operators, that maintains the expectation values of all physical
observables. It combines classical reversible logic with equally weighted
linear superposition, and is essentially free of the unitarity constraint
for quantum states. Such digital implementations can help in construction
of class P:P quantum algorithms for many linear algebra problems.

A noteworthy feature of our algorithms is that they do not make direct use
of any quantum property other than linear superposition---the constraint
of unitary evolution is reduced to an overall normalisation that can be
taken care of at the end of the computation and need not be explicitly
imposed at intermediate stages of the algorithm. Specifically, the digital
representation of quantum states makes linear algebra operations involving
action of block-diagonal Hamiltonians on a quantum state extremely simple.
The Hamiltonian blocks can be processed in superposition on a quantum
computer, while they can be handled by independent processors on a classical
parallel computer. Consequently, our algorithms can be used for classical
parallel computer simulations of quantum systems, with the same exponential
gain in temporal computational error complexity. Of course, classical and
quantum simulations will differ in the spatial resources, $N$ classical
variables vs. $\log(N)$ quantum components, but the temporal cost will be
identical with $\Delta t=\Theta(1)$. In other words, classical and quantum
complexities differ only in the cost ${\cal C}$ parametrising the resources
required to carry out a sparse matrix-vector product, and the exponential
gain in quantum spatial complexity simply arises when this product can be
evaluated using superposition of an exponentially large number of blocks.

\appendix
\section{Some Identities for Projection Operators}

Let $\{P_i\equiv|e_i\rangle\langle e_i|\}$ be a set of normalised
but not necessarily orthogonal projection operators:
\begin{equation}
P_i^2 = P_i = P_i^\dagger ~,~~
Tr(P_i P_j) = |\langle e_i|e_j\rangle|^2 \equiv |\lambda_{ij}|^2 ~.
\end{equation}
Functions of a single projection operator are linear. For instance,
the projection operators are easily exponentiated as
\begin{equation}
\exp(i\phi P_i) = 1 + (e^{i\phi}-1) P_i ~.
\end{equation}
Furthermore, functions of two projection operators reduce to quadratic
forms (note that $P_i P_j P_i = |\lambda_{ij}|^2 P_i$). In general, the
product of a string of projection operators reduces to an expression
where each projection operator appears no more than once, because
\begin{equation}
P_{i_1} P_{i_2} P_{i_3} \ldots P_{i_n} P_{i_1}
= \lambda_{i_1 i_2} \lambda_{i_2 i_3} \ldots \lambda_{i_{n-1} i_n} P_{i_1} ~.
\end{equation}
Such simplifications reduce any series of projection operators
to finite polynomials, and various identities follow.

\medskip\noindent
(A) The operator $(P_i-P_j)^2$ has the orthogonal eigenvectors
$|e_i\rangle\pm|e_j\rangle$, with degenerate eigenvalue $1-|\lambda_{ij}|^2$.
Also the operator $[P_i,P_j]^2$ has the same orthogonal eigenvectors,
with degenerate eigenvalue $|\lambda_{ij}|^4-|\lambda_{ij}|^2$.
These properties make both these operators proportional to identity
in the subspace spanned by $|e_i\rangle$ and $|e_j\rangle$.
In this subspace, therefore, we have $\{P_i,P_j\} = P_i+P_j-(P_i-P_j)^2
= P_i+P_j-1+|\lambda_{ij}|^2$,
and the identity
\begin{eqnarray}
\label{projprod}
e^{\pm i\pi P_i} e^{\pm i\pi P_j} &=& (1-2P_i) (1-2P_j) \nonumber\\
  &=& -1 + 2|\lambda_{ij}|^2 + 2[P_i,P_j] \nonumber\\
  &=& -\exp\Big( \frac{-2\sin^{-1}(|\lambda_{ij}|) [P_i,P_j]}
                      {\sqrt{|\lambda_{ij}|^2-|\lambda_{ij}|^4}} \Big) ~.
\end{eqnarray}
Eqs.(\ref{operG},\ref{stepG}) correspond to the special case of this
identity with $\lambda_{ij}=1/\sqrt{N}$. 

\medskip\noindent
(B) In the subspace spanned by $|e_i\rangle$ and $|e_j\rangle$, with a
phase choice that makes $\lambda_{ij} \equiv \langle e_i|e_j\rangle$ real,
$|e_i\rangle\pm|e_j\rangle$ are also the eigenvectors of $P_i+P_j$, with
eigenvalues $1\pm\lambda_{ij}$. The evolution taking $|e_i\rangle$ to
$|e_j\rangle$ can therefore be achieved as
\begin{eqnarray}
\exp(-i(P_i+P_j)T) |e_i\rangle
&=& \frac{1}{2} \Big[ e^{-iT(1+\lambda_{ij})} (|e_i\rangle+|e_j\rangle) 
                    + e^{-iT(1-\lambda_{ij})} (|e_i\rangle-|e_j\rangle) \Big]
    \nonumber\\
&=& \frac{1}{2} e^{-iT(1+\lambda_{ij})} \Big[ |e_i\rangle+|e_j\rangle 
              + e^{2iT\lambda_{ij}} (|e_i\rangle-|e_j\rangle) \Big]
    \nonumber\\
&=& -i e^{-iT} |e_j\rangle, \quad {\rm for}~T=\pi/(2\lambda_{ij}).
\end{eqnarray}
The Farhi-Gutmann search algorithm is the special case of this result
with $\lambda_{ij}=1/\sqrt{N}$. 

\medskip\noindent
(C) The unitary transformation generated by a projection operator is:
\begin{eqnarray}
\label{utransproj}
e^{i\phi P_i} X e^{-i\phi P_i}
&=& (1 + (e^{i\phi}-1) P_i) X (1 + (e^{-i\phi}-1) P_i) \nonumber\\
&=& X + i\sin\phi [P_i,X] + (\cos\phi-1)
    \big( \{P_i,X\} - 2 P_i X P_i \big) \nonumber\\
&=& X + i\sin\phi [P_i,X] + (\cos\phi-1)[P_i,[P_i,X]] ~.
\end{eqnarray}
In the Lie algebra language, the adjoint action of an operator is defined
as ${\rm ad}Y(X) \equiv [Y,X]$. So the above unitary transformation can be
also expressed as
\begin{equation}
e^{i\phi P_i} X e^{-i\phi P_i} = e^{i\phi ({\rm ad}P_i)} (X)
  = \Big( 1 + i\sin\phi ({\rm ad}P_i) + (\cos\phi-1)({\rm ad}P_i)^2 \Big) (X) ~.
\end{equation}
These expressions can also be derived using the identity
\begin{equation}
\label{adPident}
[P_i,[P_i,[P_i,X]]] = [P_i,X] ~~\Longleftrightarrow~~
({\rm ad}P_i)^3 (X) = {\rm ad}P_i (X) ~.
\end{equation}
When $X$ is also a projection operator, further simplification is possible
using
\begin{equation}
[P_i,[P_i,P_j]] = P_i + P_j - 1 + |\lambda_{ij}|^2 (1-2P_i),
\end{equation}
in the subspace spanned by $|e_i\rangle$ and $|e_j\rangle$.

\medskip\noindent
(D) When $X$ is a differential operator, interpreting $df/dx \equiv [d/dx,f]$,
similar algebra yields
\begin{eqnarray}
e^{i\phi P_i} \frac{d(e^{-i\phi P_i})}{dx}
&\equiv& e^{i\phi P_i} \Big[ \frac{d}{dx},e^{-i\phi P_i} \Big] \nonumber\\
%   =  e^{i\phi P_i} \frac{d}{dx} e^{-i\phi P_i} - \frac{d}{dx} \nonumber\\
&=& -iP_i \frac{d\phi}{dx} + (e^{-i\phi}-1) \frac{dP_i}{dx}
 + (2-2\cos\phi) P_i \frac{dP_i}{dx} \nonumber\\
&=& -iP_i\frac{d\phi}{dx} -i\sin\phi\frac{dP_i}{dx}
 + (1-\cos\phi) \Big[ P_i,\frac{dP_i}{dx} \Big] ~.
\end{eqnarray}
Note that $P_i^2=P_i$ leads to
\begin{equation}
P_i \frac{dP_i}{dx} + \frac{dP_i}{dx} P_i = \frac{dP_i}{dx} ~,~~
\Big[ P_i, \Big[ P_i, \frac{dP_i}{dx} \Big] \Big] = \frac{dP_i}{dx} ~.
\end{equation}

\medskip\noindent
(E) The general BCH expansion for combining exponentials of non-commuting
operators is an infinite series of nested commutators, and is cumbersome
to write down at high orders. But it can be expressed in a compact form
using exponentials of adjoint action of the operators \cite{BCHreview}:
\begin{equation}
e^A e^B = \exp\left[ A + B - \int_0^1 ds \sum_{n=1}^\infty
        \frac{(1-e^{{\rm ad}A}e^{s~{\rm ad}B})^n}{n(n+1)} B \right] ~.
\end{equation}
When the operators involved are projection operators, the identities in (C),
(D) convert exponentials of adjoint action of the operators to quadratic
polynomials, effectively summing up infinite series.

Similar simplification is possible for reflection operators as well, due to
the identity $R_i=1-2P_i$, e.g. $({\rm ad}R_i)^3 (X) = 4~{\rm ad}R_i (X)$.
The resultant reorganised formula, if necessary with appropriate truncation,
can be used as an efficient evolution operator replacing the Lie-Trotter
formula. The series in Eq.(\ref{twoprojBCH}) is an example of such a
reorganised BCH expansion.


\begin{thebibliography}{99}
\bibitem{feynman}{R.P. Feynman,
        Simulating physics with computers,
        {\it Int. J. Theor. Phys.} {\bf 21} (1982) 467-488.}
\bibitem{lloyd}{S. Lloyd,
        Universal quantum simulators,
        {\it Science} {\bf 273} (1996) 1073-1078.}
\bibitem{aharonov}{D. Aharonov and A. Ta-Shma,
        Adiabatic quantum state generation and statistical zero knowledge,
        in {\it Proc. 35th Annual ACM Symp. on Theory of Computing, STOC'03},
        San Diego, CA, 9-11 June (ACM, New York, 2003), pp.20-29.}
        %{\tt arXiv:quant-ph/0301023}
\bibitem{BACS}{D.W. Berry, G. Ahokas, R. Cleve and B.C. Sanders,
        Efficient quantum algorithms for simulating sparse Hamiltonians,
        {\it Comm. Math. Phys.} {\bf 270} (2007) 359-371.}
        % {\tt arXiv:quant-ph/0508139}
\bibitem{WBHS}{N. Wiebe, D. Berry, P. H{\o}yer and B.C. Sanders,
        Higher order decompositions of ordered operator exponentials,
        {\it J. Phys. A: Math. Theor.} {\bf 43} (2010) 065203.}
\bibitem{childs}{A.M. Childs and R. Kothari,
        Simulating sparse Hamiltonians with star decompositions,
        in {\it Theory of Quantum Computation, Communication and
        Cryptography (TQC 2010)}, Lecture Notes in Computer Science 6519
        (Springer, 2011), pp.94-103.}
        % {\tt arXiv:1003.3683}
\bibitem{suzuki}{M. Suzuki,
        Generalized Trotter's formula and systematic approximants of
        exponential operators and inner derivations with applications to
        many-body problems,
        {\it Comm. Math. Phys.} {\bf 51} (1976) 183-190.}
\bibitem{BCCKS1}{D.W. Berry, A.M. Childs, R. Cleve, R. Kothari and R.D. Somma,
        Exponential improvement in precision for simulating sparse
        Hamiltonians,
        in {\it Proc. 46th Annual ACM Symp. on Theory of Computing, STOC'14},
        New York, NY, 31 May-3 June (ACM, New York, 2014), pp.283-292.}
        % {\tt arXiv:1312.1414}
\bibitem{brent}{R.P. Brent and P. Zimmermann,
        {\it Modern Computer Arithmetic},
        (Cambridge University Press, 2010).}
\bibitem{prevcomp1}{S.A. Fenner,
        An Intuitive Hamiltonian for Quantum Search,
        arXiv:quant-ph/0004091 (2000).}
\bibitem{prevcomp2}{J. Roland and N.J. Cerf, 
        Quantum-circuit model of Hamiltonian search algorithms,
        {\it Phys. Rev. A} {\bf 68} (2003) 062311.}
        % {\tt arXiv:quant-ph/0302138}
\bibitem{LAT2014}{A. Patel, Optimisation of quantum evolution algorithms,
        in {\it Proc. 32nd International Symposium on Lattice Field Theory},
        New York, NY, 23-28 June 2014, PoS(LATTICE2014)324.}
        % {\tt arXiv:1503.01429}
\bibitem{harrow}{A.W. Harrow, A. Hassidim and S. Lloyd,
        Quantum algorithm for solving linear systems of equations,
        {\it Phys. Rev. Lett.} {\bf 103} (2009) 150502.}
        % {\tt arXiv:0811.3171}
\bibitem{clader}{B.D. Clader, B.C. Jacobs and C.R. Sprouse,
        Preconditioned quantum linear system algorithm,
        {\it Phys. Rev. Lett.} {\bf 110} (2013) 250504.}
        % {\tt arXiv:1301.2340}
%       Erratum, {\it Phys. Rev. Lett.} {\bf 111} (2013) 049903.
\bibitem{BerryChilds}{D.W. Berry and A.M. Childs,
        Black-box Hamiltonian simulation and unitary implementation,
        {\it Quant. Info. Comput.} {\bf 12} (2012) 29-62.}
\bibitem{deraedt}{H. De Raedt,
        Product formula algorithms for solving the time-dependent
        Schr\"odinger equation,
        {\it Comp. Phys. Rep.} {\bf 7} (1987) 1-72.}
\bibitem{richardson}{J.L. Richardson,
        Visualizing quantum scattering on the CM-2 supercomputer,
        {\it Comp. Phys. Comm.} {\bf 63} (1991) 84-94.}
\bibitem{trotter}{L.K. Grover,
        From Schr\"odinger's equation to the quantum search algorithm,
        {\it Pramana} {\bf 56} (2001) 333-348.}
        % {\tt arXiv:quant-ph/0109116}
\bibitem{farhi}{E. Farhi and S. Gutmann,
        Analog analogue of a digital quantum computation,
        {\it Phys. Rev. A} {\bf 57} (1998) 2403-2406.}
        % {\tt arXiv:quant-ph/9612026}
\bibitem{grover}{L.K. Grover,
        A fast quantum mechanical algorithm for database search, in
        {\it Proc. 28th Annual ACM Symposium on Theory of Computing, STOC'96},
        Philadelphia, PA, 22-24 May (ACM, New York, 1996), pp.212-219.}
        % {\tt arXiv:quant-ph/9605043}
\bibitem{BCHreview}{R. Achilles and A. Bonfiglioli,
        The early proofs of the theorem of Campbell, Baker, Hausdorff,
        and Dynkin,
        {\it Arch. Hist. Exact Sci.} {\bf 66} (2012) 295-358.}
\bibitem{nielsen}{M.A. Nielsen and I.L. Chuang,
        {\it Quantum Computation and Quantum Information},
        (Cambridge University Press, 2000), Section 6.2.}
\bibitem{TalEzer}{H. Tal-Ezer and R. Kosloff,
        An accurate and efficient scheme for propagating the time dependent
        Schr\"odinger equation,
        {\it J. Chem. Phys.} {\bf 81} (1984) 3967-3971.}
\bibitem{BCCKS2}{D.W. Berry, A.M. Childs, R. Cleve, R. Kothari and R.D. Somma,
        Simulating Hamiltonian dynamics with a truncated Taylor series,
        {\it Phys. Rev. Lett.} {\bf 114} (2015) 090502.}
        % {\tt arXiv:1412.4687}
\bibitem{Arfken}{G.B. Arfken, H.J. Weber and F.E. Harris,
        {\it Mathematical Methods for Physicists: A Comprehensive Guide},
        Seventh Edition, (Academic Press, 2011), Chapter 18.4.}
\bibitem{AMS55}{M. Abramowitz and I.A. Stegun (eds.),
        {Handbook of Mathematical Functions: With Formulas, Graphs and
        Mathematical Tables}, (Dover Publications, 1965), Chapter 9.12.}
\end{thebibliography}
\end{document}